\documentclass[10pt]{article}
\usepackage{graphicx}
\usepackage{amsmath}
\usepackage{amssymb}
\usepackage{caption2}
\begin{document}
\bibliographystyle{prsty}
\begin{center}
{\large {\bf \sc{  Two-particle  contributions and nonlocal effects  in the QCD sum rules for the axialvector tetraquark  candidate $Z_c(3900)$ }}} \\[2mm]
Zhi-Gang Wang \footnote{E-mail: zgwang@aliyun.com.  }    \\
 Department of Physics, North China Electric Power University, Baoding 071003, P. R. China
\end{center}

\begin{abstract}
In this article, we study the $Z_c(3900)$ with the QCD sum rules in details by including the two-particle scattering state contributions   and nonlocal effects between the diquark and antidiquark constituents. The two-particle scattering state contributions cannot saturate  the QCD sum rules at the hadron side,  the contribution of the $Z_c(3900)$ plays an un-substitutable role, we can saturate the QCD sum rules with or without the two-particle scattering state contributions.
If there exists a repulsive barrier or spatial distance between the diquark and antidiquark constituents,  the Feynman diagrams can be divided into the factorizable and nonfactorizable diagrams. The factorizable diagrams  consist of two colored  clusters and lead to a stable tetraquark state. The nonfactorizable Feynman diagrams correspond to the tunnelling effects, which play a minor important role in the QCD sum rules, and are consistent with the small width of the $Z_c(3900)$.
It is feasible to apply the QCD sum rules to study the tetraquark states, which  begin to receive contributions at the order   $\mathcal{O}(\alpha_s^0)$, not at the order $\mathcal{O}(\alpha_s^2)$.
\end{abstract}

PACS number: 12.39.Mk, 12.38.Lg

Key words: Tetraquark  state, QCD sum rules

\section{Introduction}

 In 2013, the BESIII collaboration  observed the  $Z_c(3900)$ in the $\pi^\pm J/\psi$ mass spectrum with a mass of $(3899.0\pm 3.6\pm 4.9)\,\rm{ MeV}$
and a width of $(46\pm 10\pm 20) \,\rm{MeV}$, respectively \cite{BES3900}. The $Z_c(3900)$ was also observed by the Belle collaboration \cite{Belle3900} and  was confirmed by the  CLEO collaboration \cite{CLEO3900}.
The average values of the mass and width of the $Z_c(3900)$ listed in {\it The Review of Particle Physics} are
$M=(3887.2\pm2.3)\,\rm{ MeV}$ and  $\Gamma=(28.2\pm 2.6) \,\rm{MeV}$, respectively \cite{PDG}.
There have been several possible  assignments for the $Z_c(3900)$,
 such as tetraquark state \cite{Maiani1303,Tetraquark3900,WangHuangTao-3900,Nielsen3900}, tetraquark molecular state \cite{Molecular3900,Molecular3900-ZhangJR,WangMolecule-3900},  hadro-charmonium \cite{hadro-charmonium-3900}, rescattering effect \cite{FSI3900}.
In 2017, the BESIII collaboration determined the spin and parity of the $Z_c(3900)$ state to be
 $J^P = 1^+$ \cite{BES-Zc3900-JP}.

In Ref.\cite{WangHuangTao-3900},  we  study the  axialvector hidden-charm  tetraquark states with the QCD sum rules,
and explore the energy scale dependence of the QCD sum rules for the tetraquark states in details for the first time. The calculations support assigning
the $X(3872)$ and $Z_c(3900)$  to be the $J^{PC}=1^{++}$ and $1^{+-}$ diquark-antidiquark type tetraquark states, respectively. In Refs.\cite{Nielsen3900,Zc3900-decay-Azizi,WangZhang-Solid}, the two-body strong decays of the $Z_c(3900)$ are studied with the QCD sum rules, which also support assigning the  $Z_c(3900)$  to be the $J^{PC}=1^{+-}$ diquark-antidiquark type tetraquark state.
In Ref.\cite{Wang-Hidden-charm}, we take the diquark and antidiquark operators as the basic constituents to construct
  the scalar, axialvector and tensor  currents to study the  mass spectrum of the ground state hidden-charm tetraquark states  with
the QCD sum rules in a comprehensive way, and  revisit the assignments of  the $X(3860)$, $X(3872)$, $X(3915)$,  $X(3940)$, $X(4160)$, $Z_c(3900)$, $Z_c(4020)$, $Z_c(4050)$, $Z_c(4055)$, $Z_c(4100)$, $Z_c(4200)$, $Z_c(4250)$, $Z_c(4430)$, $Z_c(4600)$, etc.

In those studies \cite{WangHuangTao-3900,Nielsen3900,Zc3900-decay-Azizi,WangZhang-Solid,Wang-Hidden-charm}, the axialvector current $J_\mu(x)$,
\begin{eqnarray}
J_{\mu}(x)&=&\frac{\varepsilon^{ijk}\varepsilon^{imn}}{\sqrt{2}}\Big\{u^{T}_j(x)C\gamma_5 c_k(x) \bar{d}_m(x)\gamma_\mu C \bar{c}^{T}_n(x)-u^{T}_j(x)C\gamma_\mu c_k(x)\bar{d}_m(x)\gamma_5 C \bar{c}^{T}_n(x) \Big\} \, ,
\end{eqnarray}
is chosen to interpolate the $Z_c(3900)$, where the $i$, $j$, $k$, $m$, $n$ are color indices.  The current $J_\mu(x)$ has the quantum numbers $J^{PC}=1^{+-}$,
the quantum field theory does not forbid the couplings to the two-particle scattering states $J/\psi \pi^+$, $\eta_c\rho^+$, $(D^*\bar{D}^*)^+$, $\cdots$ with the $J^{PC}=1^{+-}$. Up to now,  the contributions of the two-particle scattering states in the QCD sum rules for the hidden-charm tetraquark states have not been studied quantitatively. On the other hand, there maybe exist a repulsive barrier or spatial distance between the diquark and antidiquark constituents \cite{Wilczek-diquark,Polosa-diquark,Maiani-1903,Brodsky-PRL}, the nonlocal effects have not been taken into account in the QCD sum rules for the tetraquark states yet.
In this article, we study the $Z_c(3900)$ with the QCD sum rules in details by including the contributions of the two-particle scattering states and nonlocal effects between the diquark and antidiquark constituents, the conclusion is expected to apply to  other diquark-antidiquark type tetraquark states.

 The article is arranged as follows:  in Sect.2, we obtain  the QCD sum rules by including the contributions of the two-particle scattering states and nonlocal effects between the diquark and antidiquark constituents;    in Sect.3, we present the numerical results and discussions; and Sect.4 is reserved for our
conclusion.

\section{Two-particle  contributions and nonlocal effects in the  QCD sum rules}
Firstly, let us write down  the two-point correlation function $\Pi_{\mu\nu}(p)$,
\begin{eqnarray}
\Pi_{\mu\nu}(p)&=&i\int d^4x e^{ip \cdot x} \langle0|T\left\{J_\mu(x,\epsilon)J^{\dagger}_\nu(0,\epsilon)\right\}|0\rangle \, ,
\end{eqnarray}
where
\begin{eqnarray}
J_{\mu}(x,\epsilon)&=&\frac{\varepsilon^{ijk}\varepsilon^{imn}}{\sqrt{2}}\Big\{\bar{d}_m(x+\epsilon)\gamma_\mu C \bar{c}^{T}_n(x+\epsilon) \left[ x+\epsilon,x\right]^2 u^{T}_j(x)C\gamma_5 c_k(x)\nonumber\\
&&-\bar{d}_m(x+\epsilon)\gamma_5 C \bar{c}^{T}_n(x+\epsilon)\left[ x+\epsilon,x\right]^2 u^{T}_j(x)C\gamma_\mu c_k(x) \Big\} \, ,
\end{eqnarray}
the $i$, $j$, $k$, $m$, $n$ are color indices. The non-local current $J_\mu(x,\epsilon)$ is gauge-invariant due to the path-ordered gauge factor,
\begin{eqnarray}\label{Gauge-factor}
\left[x+\epsilon,x\right] &=&{\rm  P} \exp\left[ig_s \int_{x}^{x+\epsilon} dy^\alpha G_\alpha\left(y\right) \right]\, .
\end{eqnarray}
We choose the current $J_{\mu}(x,\epsilon)$ to interpolate the $Z_c(3900)$, and
take into account the nonlocal effects
 between the diquark and antidiquark constituents  in the current $J_\mu(x,\epsilon)$  by adding a finite four-vector $\epsilon$.
The diquark-antidiquark type tetraquark state can  be plausibly described by two diquarks in a double well potential
separated by a barrier \cite{Wilczek-diquark,Polosa-diquark}.
 At long distances, the diquark and antidiquark serve as two point color charges, and attract each other strongly.
At shorter distances,  those effects  beyond the naive one-gluon exchange force
increase at decreasing distances and produce a repulsion between
diquark and antidiquark constituents  thus form  a barrier to avoid  destroying  the diquark and antidiquark,  if  large enough \cite{Polosa-diquark}.
The diquark-antidiquark type tetraquark states emerge as QCD molecular objects made of  spatially separated colored two-quark lumps in the Born-Oppenheimer approximation \cite{Maiani-1903}.
In the dynamical picture of the tetraquark states, the large spatial distance between the diquark and antidiquark leads  to small wave-function overlap between the quark and antiquark constituents \cite{Brodsky-PRL}, which suppresses  hadronizing    to the meson-meson pairs.

\begin{figure}
 \centering
  \includegraphics[totalheight=5cm,width=10cm]{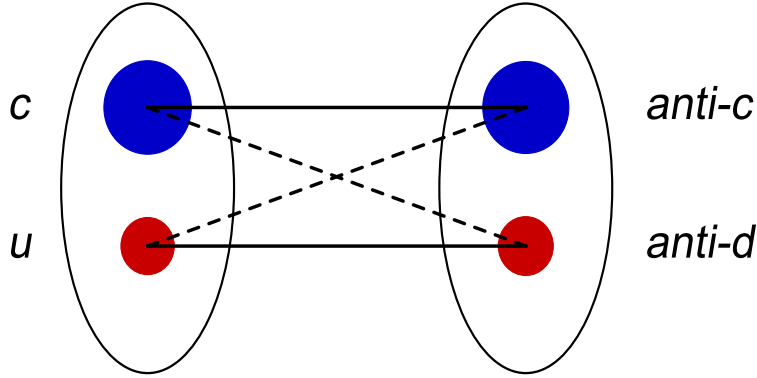}
  \caption{ The possible recombinations   of the quarks and antiquarks in the diquark ($cu$) and antidiquark ($\bar{c}\bar{d}$), which are separated by a repulsive barrier or spatial distance.   The solid lines and dashed lines represent   possible  recombinations originated from  attractions between the quarks and antiquarks in the color-singlet channels. }\label{Fierz-ccud}
\end{figure}

The finite four-vector $\epsilon^\alpha$ represents  the repulsive barrier or spatial distance between the diquark and antidiquark. For the time being, let us ignore the non-local effects of the finite four-vector $\epsilon^\alpha$  and  perform Fierz rearrangements  in the color and Dirac-spinor  spaces freely. We can arrange  the diquark-antidiquark type current $J_{\mu}(x,\varepsilon)$ into a special superposition of the     color-singlet-color-singlet type or meson-meson type currents \cite{Wang-tetra-formula,Wang-Z4020-Z4025},
\begin{eqnarray}\label{Fierz}
J_{\mu} &=&\frac{1}{2\sqrt{2}}\Big\{\,i\bar{c}i\gamma_5 c\,\bar{d}\gamma_\mu u-i\bar{c} \gamma_\mu c\,\bar{d}i\gamma_5 u+\bar{c} u\,\bar{d}\gamma_\mu\gamma_5 c
-\bar{c} \gamma_\mu \gamma_5u\,\bar{d}c  \nonumber\\
&&  - i\bar{c}\gamma^\nu\gamma_5c\, \bar{d}\sigma_{\mu\nu}u+i\bar{c}\sigma_{\mu\nu}c\, \bar{d}\gamma^\nu\gamma_5u
- i \bar{c}\sigma_{\mu\nu}\gamma_5u\,\bar{d}\gamma^\nu c+i\bar{c}\gamma^\nu u\, \bar{d}\sigma_{\mu\nu}\gamma_5c   \,\Big\} \, , \nonumber\\
&=&\frac{1}{2\sqrt{2}}\Big\{iJ^1_\mu-iJ^2_\mu- iJ^3_\mu+iJ^4_\mu+J^5_\mu-J^6_\mu    - i J^7_\mu+iJ^8_\mu   \,\Big\} \, ,
\end{eqnarray}
where
\begin{eqnarray}
J^1_\mu(x,\epsilon)&=&\bar{c}(x+\epsilon)i\gamma_5 \left[x+\epsilon,x\right]c(x)\,\bar{d}(x+\epsilon)\gamma_\mu \left[x+\epsilon,x\right]u(x)\, ,\nonumber\\
J^2_\mu(x,\epsilon)&=&\bar{c}(x+\epsilon) \gamma_\mu \left[x+\epsilon,x\right]c(x)\,\bar{d}(x+\epsilon)i\gamma_5 \left[x+\epsilon,x\right]u(x)\, ,\nonumber\\
J^3_\mu(x,\epsilon)&=&\bar{c}(x+\epsilon)\gamma^\alpha\gamma_5\left[x+\epsilon,x\right]c(x)\, \bar{d}(x+\epsilon)\sigma_{\mu\alpha}\left[x+\epsilon,x\right]u(x)\, ,\nonumber\\
J^4_\mu(x,\epsilon)&=&\bar{c}(x+\epsilon)\sigma_{\mu\alpha}\left[x+\epsilon,x\right]c(x) \,\bar{d}(x+\epsilon)\gamma^\alpha\gamma_5\left[x+\epsilon,x\right]u(x)\, ,\nonumber\\
J^5_\mu(x,\epsilon)&=&\bar{c}(x+\epsilon) \left[x+\epsilon,x\right]u(x)\,\bar{d}(x+\epsilon)\gamma_\mu\gamma_5 \left[x+\epsilon,x\right]c(x)\, ,\nonumber\\
J^6_\mu(x,\epsilon)&=&\bar{c}(x+\epsilon) \gamma_\mu \gamma_5\left[x+\epsilon,x\right]u(x)\,\bar{d}(x+\epsilon)\left[x+\epsilon,x\right]c(x)\, ,\nonumber\\
J^7_\mu(x,\epsilon)&=&\bar{c}(x+\epsilon)\sigma_{\mu\alpha}\gamma_5\left[x+\epsilon,x\right]u(x)\,\bar{d}(x+\epsilon)\gamma^\alpha \left[x+\epsilon,x\right]c(x)\, ,\nonumber\\
J^8_\mu(x,\epsilon)&=&\bar{c}(x+\epsilon)\gamma^\alpha \left[x+\epsilon,x\right]u(x) \, \bar{d}(x+\epsilon)\sigma_{\mu\alpha}\gamma_5\left[x+\epsilon,x\right]c(x)\, .
\end{eqnarray}
In Fig.\ref{Fierz-ccud}, we illustrate the Fierz rearrangements  or recombinations  diagrammatically. The currents $J_\mu^n(x,\epsilon)$ with $n=1$, $2$, $\cdots$, $8$ are  gauge-invariant due to the path-ordered gauge factor $\left[x+\epsilon, x\right] $.
In the following, we drop the gauge factors by taking  the Fock-Schwinger gauge $y_\alpha G^\alpha(y)=0$ in Eq.\eqref{Gauge-factor}.   In the QCD sum rules, it is convenient to  calculate the Wilson's coefficients in the operator product expansion by taking  the gluon field as an external field and choose the  Fock-Schwinger gauge \cite{Shifman-Fortsch}.
In fact, the repulsive barrier or spatial distance between the diquark and antidiquark pair frustrates the Fierz rearrangements or recombinations \cite{Wilczek-diquark,Polosa-diquark,Maiani-1903,Brodsky-PRL}.

We expend the currents $J_\mu(x,\epsilon)$ and  $J^n_\mu(x,\epsilon)$ with $n=1$, $2$, $\cdots$, $8$ in terms of Taylor series of $\epsilon$,
\begin{eqnarray}
J_\mu(x,\epsilon)&=&J_\mu(x,0)+ \frac{\partial J_\mu(x,\epsilon)}{\partial\epsilon^\alpha}\mid_{\epsilon=0} \epsilon^\alpha+\frac{1}{2} \frac{\partial^2 J_\mu(x,\epsilon)}{\partial\epsilon^\alpha\partial\epsilon^\beta}\mid_{\epsilon=0} \epsilon^\alpha\epsilon^\beta+\cdots\, , \nonumber\\
J_\mu^n(x,\epsilon)&=&J^n_\mu(x,0)+ \frac{\partial J^n_\mu(x,\epsilon)}{\partial\epsilon^\alpha}\mid_{\epsilon=0} \epsilon^\alpha+\frac{1}{2} \frac{\partial^2 J^n_\mu(x,\epsilon)}{\partial\epsilon^\alpha\partial\epsilon^\beta}\mid_{\epsilon=0} \epsilon^\alpha\epsilon^\beta+\cdots\, ,
\end{eqnarray}
then  expend the correlation function $\Pi_{\mu\nu}(p)$ also in terms of Taylor series of $\epsilon$,
\begin{eqnarray}
\Pi_{\mu\nu}(p)&=&\Pi_{\mu\nu}\left(\mathcal{O}(\epsilon^0)\right)+\Pi_{\mu\nu}\left(\mathcal{O}(\epsilon^1)\right)+\Pi_{\mu\nu}\left(\mathcal{O}(\epsilon^2)\right)+\cdots\, , \end{eqnarray}
where the components $\Pi_{\mu\nu}\left(\mathcal{O}(\epsilon^i)\right)$ with $i=0$, $1$, $2$, $\cdots$ represent the contributions of the order $\mathcal{O}(\epsilon^i)$.
We write the contributions in the leading order explicitly,
\begin{eqnarray}
\Pi_{\mu\nu}\left(\mathcal{O}(\epsilon^0)\right)&=&\Pi(p^2)\left(-g_{\mu\nu}+\frac{p_{\mu}p_{\nu}}{p^2} \right)+\Pi_0(p^2)\frac{p_{\mu}p_{\nu}}{p^2}\, ,
\end{eqnarray}
according to Lorentz covariance, where the $\Pi(p^2)$ and $\Pi_0(p^2)$ denote the spin $1$ and $0$ components respectively.
In this article, we study the component $\Pi(p^2)$.

In Refs.\cite{WangHuangTao-3900,Wang-Hidden-charm}, we observe that the axialvector tetraquark current $J_{\mu}(x,0)$ couples potentially to the $Z_c(3900)$,
\begin{eqnarray}
\langle 0|J_\mu(0,0)|Z_c(p)\rangle&=&\lambda_Z\,\varepsilon_\mu\, ,
\end{eqnarray}
where the $\lambda_Z$ is the current-hadron coupling constant or pole residue, the $\varepsilon_\mu$ is the polarization vector.

The color-singlet-color-singlet type currents couple potentially to the meson-meson pairs or tetraquark molecular states. In the following,
we write down the couplings to the  meson-meson pairs explicitly,
\begin{eqnarray}
\langle 0|J^1_\mu(0,0)|\eta_c(q)\rho(p-q)\rangle&=&\frac{f_{\eta_c}m^2_{\eta_c}}{2m_c}f_{\rho}m_{\rho}\varepsilon_{\mu}^\rho\, ,
\nonumber\\
\langle 0|J^2_\mu(0,0)|\pi(q)J/\psi(p-q)\rangle&=&\frac{f_{\pi}m^2_{\pi}}{m_u+m_d}f_{J/\psi}m_{J/\psi}\varepsilon_{\mu}^{J/\psi}\, , \nonumber\\
\langle 0|J^2_\mu(0,0)|\pi(q)\psi^\prime(p-q)\rangle&=&\frac{f_{\pi}m^2_{\pi}}{m_u+m_d}f_{\psi^\prime}m_{\psi^\prime}\varepsilon_{\mu}^{\psi^\prime}\, ,
\end{eqnarray}

\begin{eqnarray}
\langle 0|J^3_\mu(0,0)|\eta_c(q)\rho(p-q)\rangle&=&-if_{\eta_c}q^{\alpha}if_{\rho}^T\left[\varepsilon^\rho_\mu (p-q)_\alpha-\varepsilon^\rho_\alpha (p-q)_\mu \right]
\, ,\nonumber\\
\langle 0|J^3_\mu(0,0)|\eta_c(q)b_1(p-q)\rangle&=&-if_{\eta_c}q^{\alpha}if_{b_1}\varepsilon_{\mu\alpha\sigma\tau}\varepsilon_{h_1}^{\sigma} (p-q)^{\tau}\, ,\nonumber\\
\langle 0|J^3_\mu(0,0)|\chi_{c1}(q)\rho(p-q)\rangle&=&f_{\chi_{c1}}m_{\chi_{c1}}\varepsilon^{\alpha}_{\chi_{c1}}if^T_{\rho}\left[\varepsilon^\rho_\mu (p-q)_\alpha-\varepsilon^\rho_\alpha (p-q)_\mu \right]\, ,
\end{eqnarray}

\begin{eqnarray}
\langle 0|J^4_\mu(0,0)|\pi(q)J/\psi(p-q)\rangle&=&-if_{\pi}q^{\alpha}if_{J/\psi}^T\left[\varepsilon^{J/\psi}_\mu (p-q)_\alpha-\varepsilon^{J/\psi}_\alpha (p-q)_\mu \right]\, , \nonumber\\
\langle 0|J^4_\mu(0,0)|\pi(q)\psi^\prime(p-q)\rangle&=&-if_{\pi}q^{\alpha}if_{\psi^\prime}^T\left[\varepsilon^{\psi^\prime}_\mu (p-q)_\alpha-\varepsilon^{\psi^\prime}_\alpha (p-q)_\mu \right]\, , \nonumber\\
\langle 0|J^4_\mu(0,0)|\pi(q)h_{c}(p-q)\rangle&=&-if_{\pi}q^{\alpha}if_{h_c}\varepsilon_{\mu\alpha\sigma\tau}\varepsilon_{h_c}^{\sigma} (p-q)^{\tau}\, ,\nonumber\\
\langle 0|J^4_\mu(0,0)|\pi(q)h^\prime_{c}(p-q)\rangle&=&-if_{\pi}q^{\alpha}if_{h^\prime_c}\varepsilon_{\mu\alpha\sigma\tau}\varepsilon_{h^\prime_c}^{\sigma} (p-q)^{\tau}\, ,\nonumber\\
\langle 0|J^4_\mu(0,0)|a_{1}(q)J/\psi(p-q)\rangle&=&f_{a_{1}}m_{a_{1}}\varepsilon^{\alpha}_{a_{1}}if^T_{J/\psi}\left[\varepsilon^{J/\psi}_\mu (p-q)_\alpha-\varepsilon^{J/\psi}_\alpha (p-q)_\mu \right]\, ,
\end{eqnarray}

\begin{eqnarray}
\langle 0|J^{5/6}_\mu(0,0)|\bar{D}_0(q)D(p-q)\rangle&=&-if_{D_0}m_{D_0}f_{D} (p-q)_{\mu}\, ,\nonumber\\
\langle 0|J^{7/8}_\mu(0,0)|\bar{D}^*(q)D^*(p-q)\rangle&=&f_{D^*}m_{D^*}\varepsilon^{\alpha}_{D^*}if_{D^*}^T\varepsilon_{\mu\alpha\sigma\tau} \varepsilon^\sigma_{D^*}(p-q)^{\tau}\, ,\nonumber\\
\langle 0|J^{7/8}_\mu(0,0)|\bar{D}_0(q)D^*(p-q)\rangle&=&f_{D_0}q^{\alpha}if_{D^*}^T\varepsilon_{\mu\alpha\sigma\tau} \varepsilon^\sigma_{D^*}(p-q)^{\tau}\, ,
\end{eqnarray}
and examine whether or not  the color-singlet-color-singlet type currents can be saturated with the meson-meson pairs, where
the $\varepsilon_\mu$ are the polarization vectors of the vector and axialvector mesons, the $f_{\eta_c}$, $f_{\rho}$, $f^T_{\rho}$, $f_{\pi}$, $f_{J/\psi}$, $f_{\psi^\prime}$, $f_{J/\psi}^T$, $f_{\psi^\prime}^T$, $f_{h_c}$, $f_{h^\prime_c}$, $f_{b_1}$, $f_{\chi_{c1}}$, $f_{a_1}$, $f_{D_0}$, $f_D$, $f_{D^*}$ and $f_{D*}^T$ are the decay constants.

At the hadron side,
we take into account the contributions of the tetraquark candidate $Z_c(3900)$ and the two-particle scattering state contributions from the
$\pi J/\psi$, $\pi h_c$, $\eta_c \rho$, $D^* D^*$, $D_0 D $, $\eta_c b_1$, $\chi_{c1}\rho$, $a_1 J/\psi$, $D_0 D^*$, $\pi \psi^\prime$, $\pi h^\prime_c$, $\pi \psi^{\prime\prime}$  below the threshold of the $Z_c(4430)$,
which can be assigned to the first radial excitation of the $Z_c(3900)$ \cite{Z4430-1405,Nielsen-1401,Wang4430},
\begin{eqnarray}
\Pi(p^2)&=&\frac{\lambda_Z^2}{M_Z^2-p^2}\nonumber\\
&&+\lambda_{\eta_c \rho;11}^2\int_{m^2_{\eta_c \rho}}^{s_0}ds \frac{1}{s-p^2}\frac{\sqrt{\lambda(s,m_{\eta_c}^2,m^2_{\rho})}}{s}\left[1+\frac{\lambda(s,m_{\eta_c}^2,m^2_{\rho})}{12sm_\rho^2} \right]\nonumber\\
&&+\lambda_{\pi J/\psi;22}^2\int_{m^2_{\pi J/\psi}}^{s_0}ds \frac{1}{s-p^2}\frac{\sqrt{\lambda(s,m_{\pi}^2,m^2_{J/\psi})}}{s}\left[1+\frac{\lambda(s,m_{\pi}^2,m^2_{J/\psi})}{12sm_{J/\psi}^2} \right]\nonumber\\
&&+\frac{\lambda_{\eta_c \rho;33}^2}{4}\int_{m^2_{\eta_c \rho}}^{s_0}ds \frac{1}{s-p^2}\frac{\sqrt{\lambda(s,m_{\eta_c}^2,m^2_{\rho})}}{s}\left[(s-m_{\eta_c}^2-m^2_{\rho})^2-\frac{\lambda(s,m_{\eta_c}^2,m^2_{\rho})(s-m_\rho^2)}{3s} \right]\nonumber\\
&&+\frac{\lambda_{\pi J/\psi;44}^2}{4}\int_{m^2_{\pi J/\psi}}^{s_0}ds \frac{1}{s-p^2}\frac{\sqrt{\lambda(s,m_{\pi}^2,m^2_{J/\psi})}}{s}\left[(s-m_{\pi}^2-m^2_{J/\psi})^2-\frac{\lambda(s,m_{\pi}^2,m^2_{J/\psi})(s-m_{J/\psi}^2)}{3s} \right]\nonumber\\
&&+\lambda_{\eta_c \rho;13}^2\int_{m^2_{\eta_c \rho}}^{s_0}ds \frac{1}{s-p^2}\frac{\sqrt{\lambda(s,m_{\eta_c}^2,m^2_{\rho})}}{s}\left[\frac{\lambda(s,m_{\eta_c}^2,m^2_{\rho})}{6}-(s-m_{\eta_c}^2-m^2_{\rho}) \right]\nonumber\\
&&+\lambda_{\pi J/\psi;24}^2\int_{m^2_{\pi J/\psi}}^{s_0}ds \frac{1}{s-p^2}\frac{\sqrt{\lambda(s,m_{\pi}^2,m^2_{J/\psi})}}{s}\left[\frac{\lambda(s,m_{\pi}^2,m^2_{J/\psi})}{6}-(s-m_{\pi}^2-m^2_{J/\psi}) \right]\nonumber\\
&&+\lambda_{\eta_c b_1;33}^2\int_{m^2_{\eta_c b_1}}^{s_0}ds \frac{1}{s-p^2}\frac{\sqrt{\lambda(s,m_{\eta_c}^2,m^2_{b_1})}}{s}\frac{\lambda(s,m_{\eta_c}^2,m^2_{b_1})}{6} \nonumber\\
&&+\lambda_{\chi_{c1} \rho;33}^2\int_{m^2_{\chi_{c1}  \rho}}^{s_0}ds \frac{1}{s-p^2}\frac{\sqrt{\lambda(s,m_{\chi_{c1} }^2,m^2_{\rho})}}{s}\frac{\lambda(s,m_{\chi_{c1}}^2,m^2_{\rho})(2s+m_\rho^2+2m^2_{\chi_{c1}})}{12sm^2_{\chi_{c1}}}\nonumber\\
&&+\lambda_{\pi h_c;44}^2\int_{m^2_{\pi h_c}}^{s_0}ds \frac{1}{s-p^2}\frac{\sqrt{\lambda(s,m_{\pi}^2,m^2_{h_c})}}{s}\frac{\lambda(s,m_{\pi}^2,m^2_{h_c})}{6} \nonumber\\
&&+\lambda_{a_{1} J/\psi;44}^2\int_{m^2_{a_{1}  J/\psi}}^{s_0}ds \frac{1}{s-p^2}\frac{\sqrt{\lambda(s,m_{a_{1} }^2,m^2_{J/\psi})}}{s}\frac{\lambda(s,m_{a_{1}}^2,m^2_{J/\psi})(2s+m_{J/\psi}^2+2m^2_{a_{1}})}{12sm^2_{a_{1}}}\nonumber\\
&&+\lambda_{D_0 D;55/66}^2\int_{m^2_{D_0D}}^{s_0}ds \frac{1}{s-p^2}\frac{\sqrt{\lambda(s,m_{D_0}^2,m^2_{D})}}{s}\frac{\lambda(s,m_{D_0}^2,m^2_{D})}{12s}\nonumber\\
&&+\lambda_{D^*D^*;77/88}^2\int_{m^2_{D^*D^*}}^{s_0}ds \frac{1}{s-p^2}\frac{\sqrt{\lambda(s,m_{D^*}^2,m^2_{D^*})}}{s}\left[2m_{D^*}^2+\frac{\lambda(s,m_{D^*}^2,m^2_{D^*})(s+m_{D^*}^2)}{6sm_{D^*}^2} \right]\nonumber\\
&&+\lambda_{D_0D^*;77/88}^2\int_{m^2_{D_0D^*}}^{s_0}ds \frac{1}{s-p^2}\frac{\sqrt{\lambda(s,m_{D_0}^2,m^2_{D^*})}}{s}\frac{\lambda(s,m_{D_0}^2,m^2_{D^*})}{6}\nonumber
\end{eqnarray}
\begin{eqnarray}
&&+\lambda_{\pi \psi^\prime;22}^2\int_{m^2_{\pi \psi^\prime}}^{s_0}ds \frac{1}{s-p^2}\frac{\sqrt{\lambda(s,m_{\pi}^2,m^2_{\psi^\prime})}}{s}\left[1+\frac{\lambda(s,m_{\pi}^2,m^2_{\psi^\prime})}{12sm_{\psi^\prime}^2} \right]\nonumber\\
&&+\frac{\lambda_{\pi \psi^\prime;44}^2}{4}\int_{m^2_{\pi \psi^\prime}}^{s_0}ds \frac{1}{s-p^2}\frac{\sqrt{\lambda(s,m_{\pi}^2,m^2_{\psi^\prime})}}{s}\left[(s-m_{\pi}^2-m^2_{\psi^\prime})^2-\frac{\lambda(s,m_{\pi}^2,m^2_{\psi^\prime})(s-m_{\psi^\prime}^2)}{3s} \right]\nonumber\\
&&+\lambda_{\pi \psi^\prime;24}^2\int_{m^2_{\pi \psi^\prime}}^{s_0}ds \frac{1}{s-p^2}\frac{\sqrt{\lambda(s,m_{\pi}^2,m^2_{\psi^\prime})}}{s}\left[\frac{\lambda(s,m_{\pi}^2,m^2_{\psi^\prime})}{6}-(s-m_{\pi}^2-m^2_{\psi^\prime}) \right]\nonumber\\
&&+\lambda_{\pi h_c;44}^2\int_{m^2_{\pi h^\prime_c}}^{s_0}ds \frac{1}{s-p^2}\frac{\sqrt{\lambda(s,m_{\pi}^2,m^2_{h^\prime_c})}}{s}\frac{\lambda(s,m_{\pi}^2,m^2_{h^\prime_c})}{6} \nonumber\\
&&+\left(\pi \psi^\prime \to \pi \psi^{\prime\prime}  \right)+\cdots\, ,
\end{eqnarray}
where
\begin{eqnarray}
\lambda_{\eta_c \rho;11}^2&=&\frac{1}{128\pi^2}\frac{f_{\eta_c}^2m_{\eta_c}^4f_\rho^2m_\rho^2}{4m_c^2}\, ,\nonumber\\
\lambda_{\pi J/\psi;22}^2&=&\frac{1}{128\pi^2}\frac{f_{\pi}^2m_{\pi}^4f_{J/\psi}^2m_{J/\psi}^2}{(m_u+m_d)^2}\, ,
\end{eqnarray}
\begin{eqnarray}
\lambda_{\eta_c \rho;33}^2&=&\frac{1}{128\pi^2}f_{\eta_c}^2f_\rho^{T2}\, ,\nonumber\\
\lambda_{\eta_c b_1;33}^2&=&\frac{1}{128\pi^2}f_{\eta_c}^2f_{b_1}^{2}\, ,\nonumber\\
\lambda_{\chi_{c1} \rho;33}^2&=&\frac{1}{128\pi^2}f_{\chi_{c1}}^2m_{\chi_{c1}}^2f_{\rho}^{T2}\, ,
\end{eqnarray}
\begin{eqnarray}
\lambda_{\pi J/\psi;44}^2&=&\frac{1}{128\pi^2}f_{\pi}^2f_{J/\psi}^{T2}\, ,\nonumber\\
\lambda_{\pi h_c;44}^2&=&\frac{1}{128\pi^2}f_{\pi}^2f_{h_c}^{2}\, ,\nonumber\\
\lambda_{a_{1} J/\psi;44}^2&=&\frac{1}{128\pi^2}f_{a_{1}}^2m_{a_{1}}^2f_{J/\psi}^{T2}\, ,
\end{eqnarray}
\begin{eqnarray}
\lambda_{\eta_c \rho;13}^2&=&\frac{1}{128\pi^2}\frac{f_{\eta_c}^2m_{\eta_c}^2f_\rho m_\rho f_{\rho}^T}{2m_c}\, ,\nonumber\\
\lambda_{\pi J/\psi;24}^2&=&\frac{1}{128\pi^2}\frac{f_{\pi}^2m_{\pi}^2f_{J/\psi} m_{J/\psi} f_{J/\psi}^T}{m_u+m_d}\, ,
\end{eqnarray}
\begin{eqnarray}
\lambda_{D_0 D;55/66}^2&=&\frac{1}{128\pi^2}f_{D_0}^2m_{D_0}^2f_{D}^{2}\, ,\nonumber\\
\lambda_{D^* D^*;77/88}^2&=&\frac{1}{128\pi^2}f_{D^*}^2m_{D^*}^2f_{D^*}^{T2}\, ,\nonumber\\
\lambda_{D_0 D^*;77/88}^2&=&\frac{1}{128\pi^2}f_{D_0}^2f_{D^*}^{T2}\, ,
\end{eqnarray}

\begin{eqnarray}
\lambda_{\psi^\prime;22}^2&=&\frac{1}{128\pi^2}\frac{f_{\pi}^2m_{\pi}^4f_{\psi^\prime}^2m_{\psi^\prime}^2}{(m_u+m_d)^2}\, , \nonumber\\
\lambda_{\pi \psi^\prime;44}^2&=&\frac{1}{128\pi^2}f_{\pi}^2f_{\psi^\prime}^{T2}\, ,\nonumber\\
\lambda_{\pi \psi^\prime;24}^2&=&\frac{1}{128\pi^2}\frac{f_{\pi}^2m_{\pi}^2f_{\psi^\prime} m_{\psi^\prime} f_{\psi^\prime}^T}{m_u+m_d}\, ,\nonumber\\
\lambda_{\pi h^\prime_c;44}^2&=&\frac{1}{128\pi^2}f_{\pi}^2f_{h^\prime_c}^{2}\, ,
\end{eqnarray}
$\lambda(a,b,c)=a^2+b^2+c^2-2ab-2ac-2bc$, $m^2_{\eta_c \rho}=(m_{\eta_c}+m_\rho)^2$, $m^2_{\pi J/\psi}=(m_{\pi}+m_{J/\psi})^2$, $m^2_{\eta_c b_1}=(m_{\eta_c}+m_{b_1})^2$,
$m^2_{\pi h_c}=(m_{\pi}+m_{h_c})^2$, $m^2_{\chi_{c1} \rho}=(m_{\chi_{c1}}+m_{\rho})^2$, $m^2_{a_1 J/\psi}=(m_{a_1}+m_{J/\psi})^2$,
$m^2_{D_0 D}=(m_{D_0}+m_{D})^2$, $m^2_{D^* D^*}=(m_{D^*}+m_{D^*})^2$, $m^2_{D_0 D^*}=(m_{D_0}+m_{D^*})^2$,
$m^2_{\pi \psi^\prime}=(m_{\pi}+m_{\psi^\prime})^2$,
$m^2_{\pi h^\prime_c}=(m_{\pi}+m_{h^\prime_c})^2$, the continuum threshold parameter $\sqrt{s_0}\leq M_{Z_c(4430)}$. In Table \ref{meson-meson-pair}, we present the
thresholds for the relevant meson-meson pairs from the Particle Data Group \cite{PDG}.

\begin{table}
\begin{center}
\begin{tabular}{|c|c|c|c|c|c|c|c|c|c|c|c|c|}\hline\hline

                  &$\pi J/\psi$  &$\pi h_c$  &$\eta_c \rho$  &$D^* D^*$  &$D_0 D$  &$\eta_c b_1$  &$\chi_{c1}\rho$  &$a_1 J/\psi$ &$D_0 D^*$  \\ \hline

Thresholds (GeV)  &$3.24$        &$3.67$     &$3.76$         &$4.02$     &$4.19$   &$4.21$        &$4.29$           &$4.33$       &$4.33$   \\ \hline

                  &$\pi\psi^\prime$ &$\pi\psi^{\prime\prime}$ &$\pi h^\prime_c$       &           &         &           &           &             &   \\ \hline

Thresholds (GeV)  &$3.83$           &$4.12$                   &$4.10$                 &           &         &           &           &             &   \\ \hline\hline

\end{tabular}
\end{center}
\caption{ The thresholds for the relevant meson-meson pairs from the Particle Data Group. }\label{meson-meson-pair}
\end{table}

\begin{figure}
 \centering
  \includegraphics[totalheight=4cm,width=10cm]{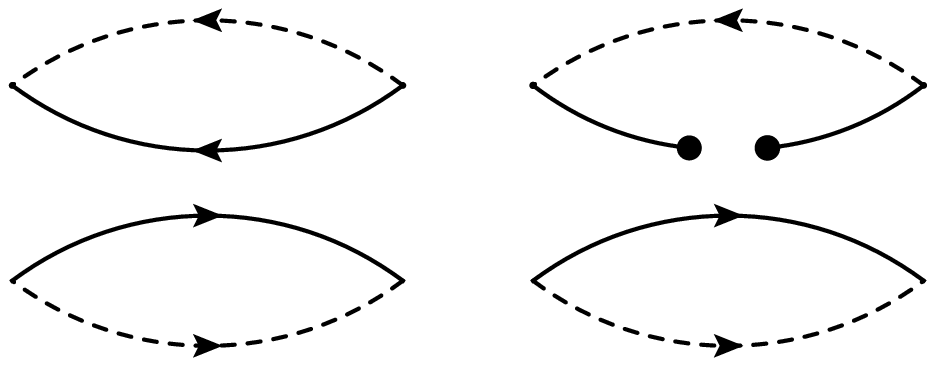}\\
    \vspace{1cm}
  \includegraphics[totalheight=4cm,width=10cm]{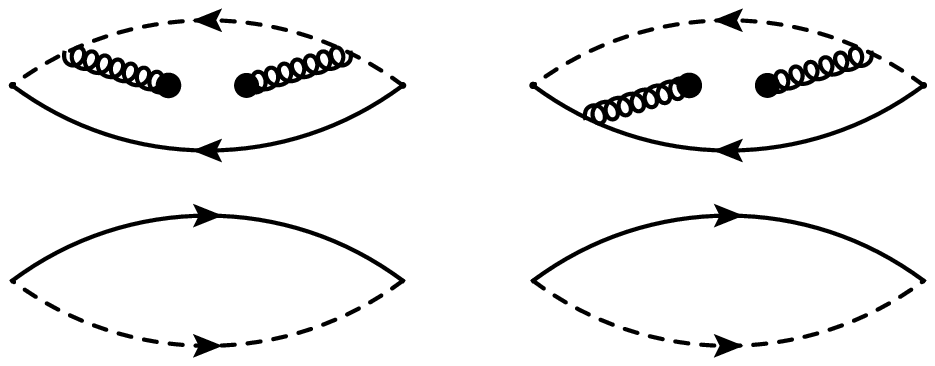}\\
  \vspace{1cm}
  \includegraphics[totalheight=4cm,width=10cm]{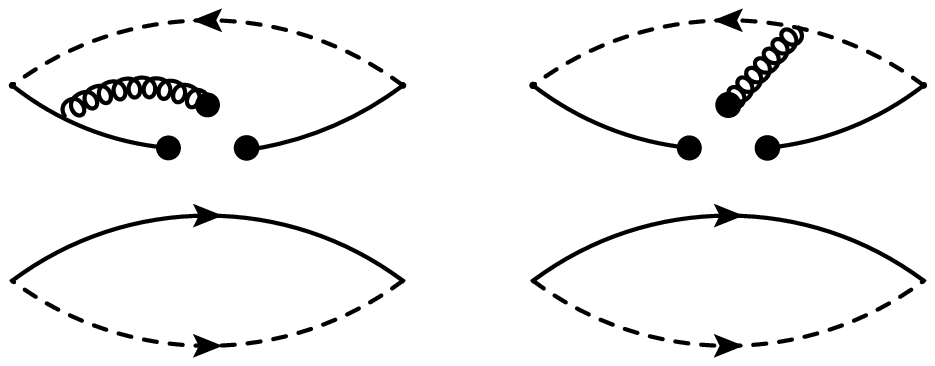}\\
\vspace{1cm}
 \includegraphics[totalheight=4cm,width=10cm]{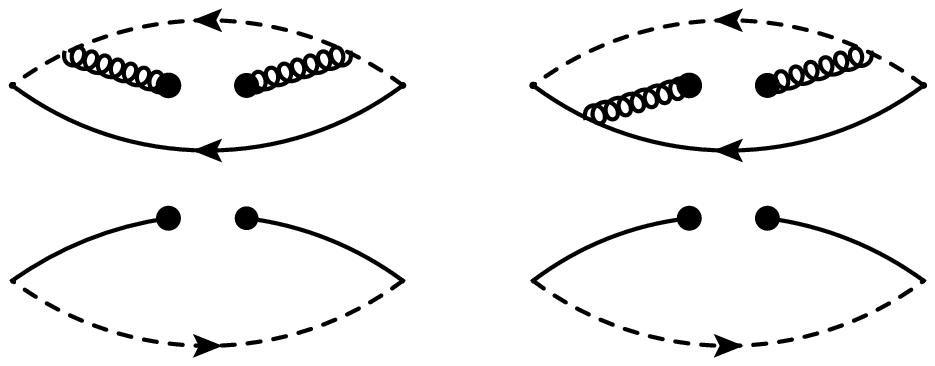}
  \caption{ The factorizable Feynman diagrams contribute  to the perturbative term, $\langle\bar{q}q\rangle$, $\langle\frac{\alpha_sGG}{\pi}\rangle$, $\langle\bar{q}g_s\sigma Gq\rangle$ and $\langle\bar{q}q\rangle\langle\frac{\alpha_sGG}{\pi}\rangle$,  where the solid lines and dashed lines denote the light quarks and heavy quarks, respectively. Other diagrams obtained by interchanging of the  light quark lines and heavy quark lines are implied. }\label{Lowest-diagram}
\end{figure}

\begin{figure}
 \centering
  \includegraphics[totalheight=4cm,width=10cm]{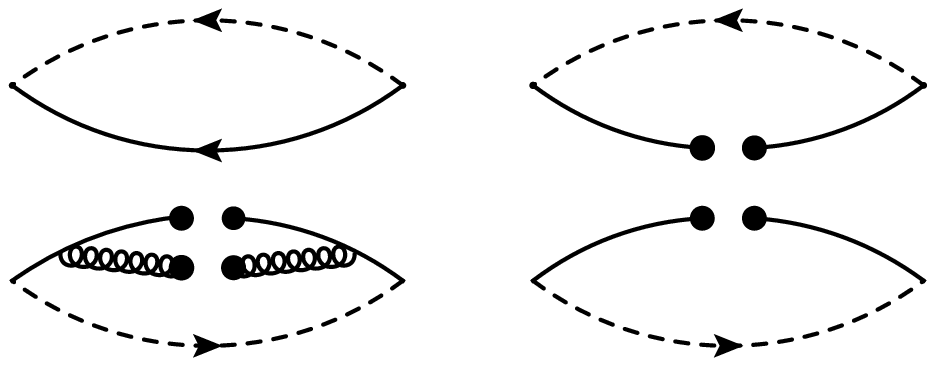}\\
 \vspace{1cm}
\includegraphics[totalheight=4cm,width=10cm]{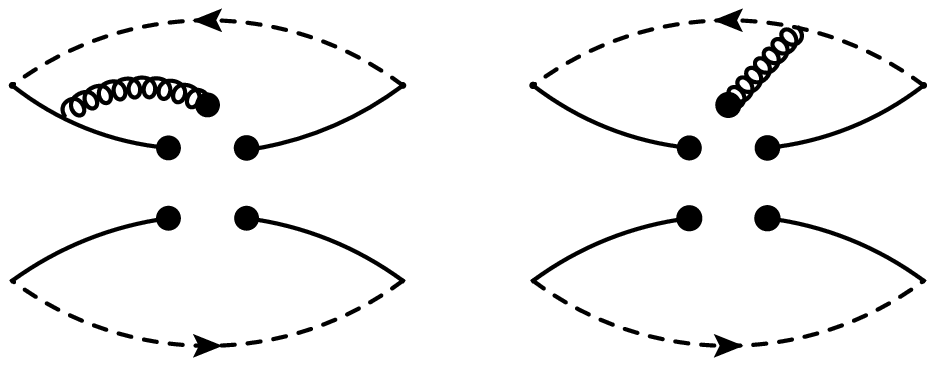}\\
\vspace{1cm}
 \includegraphics[totalheight=4cm,width=10cm]{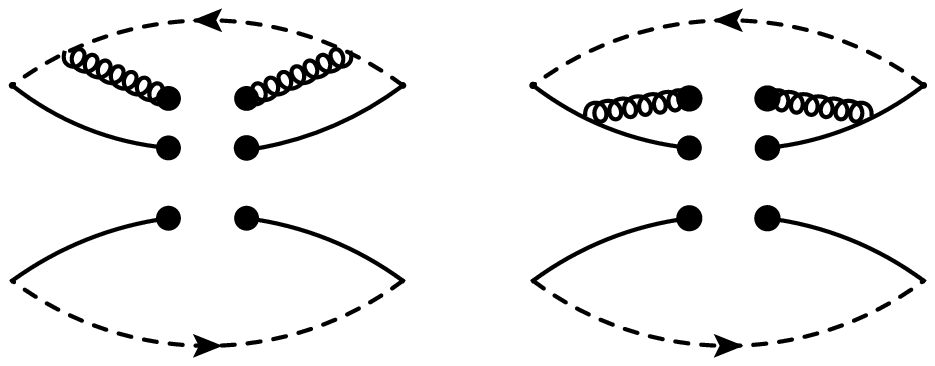}\\
 \vspace{1cm}
  \includegraphics[totalheight=4cm,width=10cm]{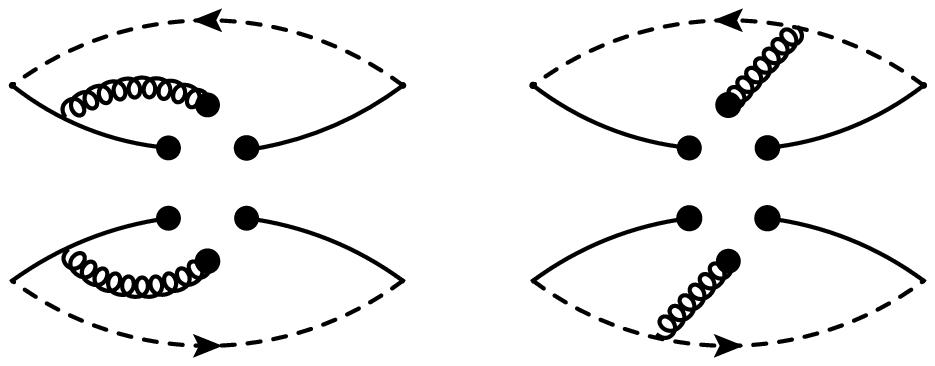}
 \caption{ The  factorizable Feynman diagrams  contribute  to  the $\langle\bar{q}q\rangle\langle\frac{\alpha_sGG}{\pi}\rangle$,  $\langle\bar{q}q\rangle^2$,  $\langle\bar{q}q\rangle\langle\bar{q}g_s\sigma Gq\rangle$, $\langle\bar{q}q\rangle^2\langle\frac{\alpha_sGG}{\pi}\rangle$ and $\langle\bar{q}g_s\sigma Gq\rangle^2$. Other diagrams obtained by interchanging of the light quark lines and  heavy quark lines are implied.}\label{fac-qq-qqg}
\end{figure}

\begin{figure}
 \centering
    \includegraphics[totalheight=4cm,width=5cm]{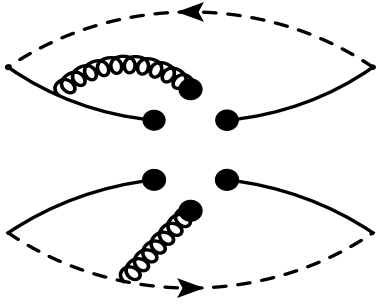}
 \caption{ The  factorizable Feynman diagrams  contribute  to  the $\langle\bar{q}g_s\sigma Gq\rangle^2$. Other diagrams obtained by interchanging of the light quark lines and  heavy quark lines are implied.}\label{fac-qqg-qqg}
\end{figure}

\begin{figure}
 \centering
    \includegraphics[totalheight=4cm,width=10cm]{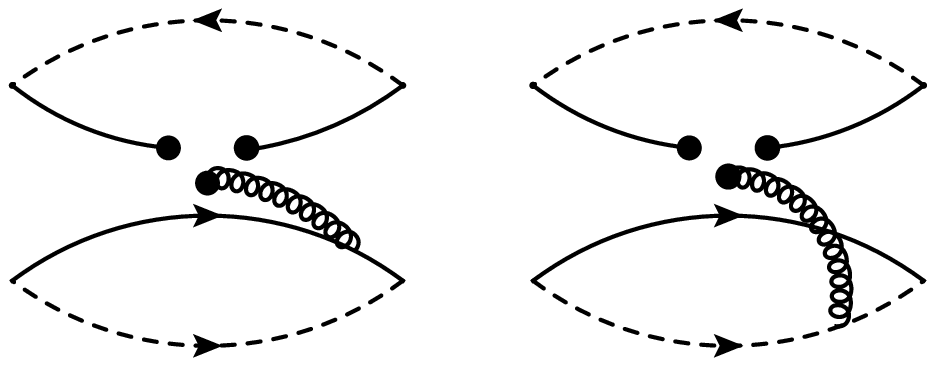}\\
     \vspace{1cm}
    \includegraphics[totalheight=4cm,width=10cm]{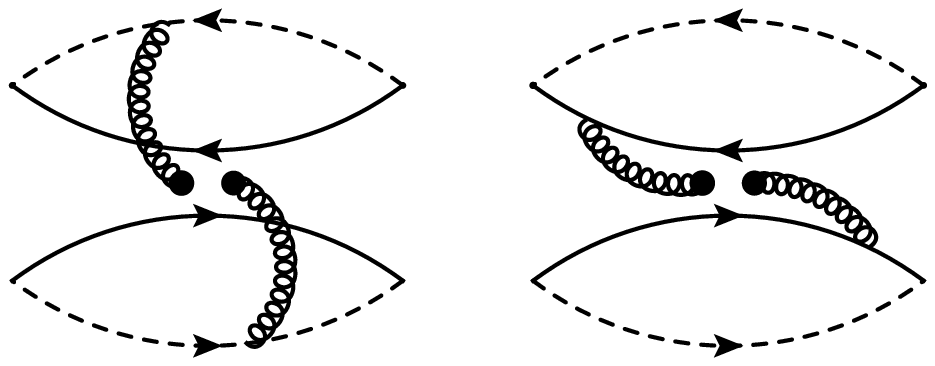}\\
 \vspace{1cm}
  \includegraphics[totalheight=4cm,width=10cm]{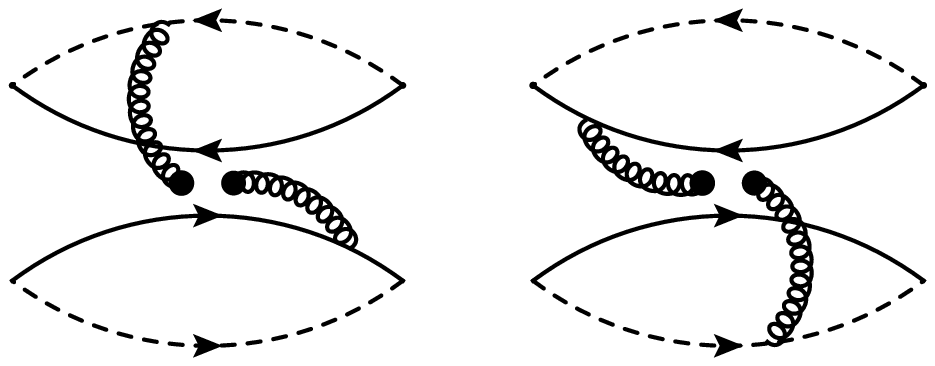}\\
  \vspace{1cm}
  \includegraphics[totalheight=4cm,width=10cm]{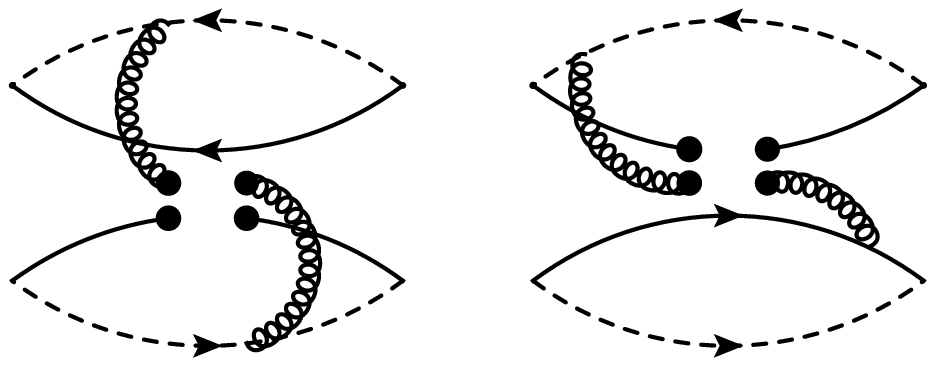}
 \caption{ The  nonfactorizable Feynman diagrams  contribute  to  the  $\langle\bar{q}g_s\sigma Gq\rangle$, $\langle\frac{\alpha_sGG}{\pi}\rangle$ and $\langle\bar{q}q\rangle\langle\frac{\alpha_sGG}{\pi}\rangle$. Other diagrams obtained by interchanging of the  light quark lines  and heavy quark lines are implied.}\label{nonfac-qqg}
\end{figure}

\begin{figure}
 \centering
 \includegraphics[totalheight=4cm,width=10cm]{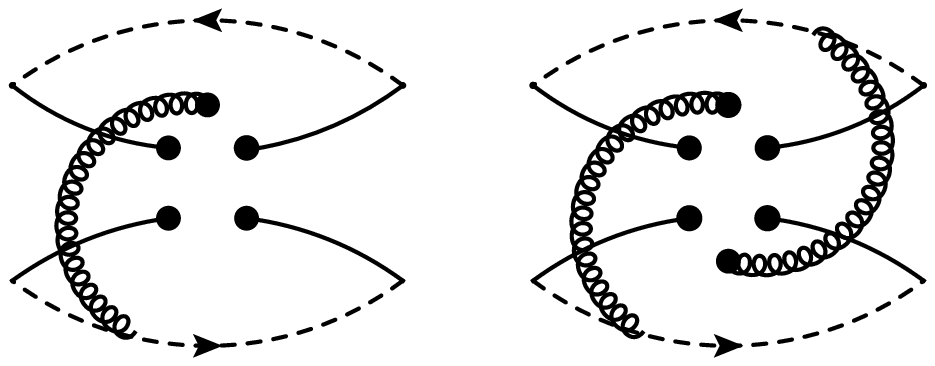}\\
 \vspace{1cm}
  \includegraphics[totalheight=4cm,width=4.5cm]{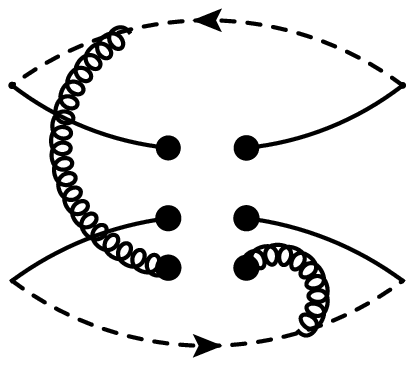}
 \caption{ The  nonfactorizable Feynman diagrams  contribute  to  the $\langle\bar{q}q\rangle\langle\bar{q}g_s\sigma Gq\rangle$, $\langle\bar{q}g_s\sigma Gq\rangle^2$ and $\langle\bar{q}q\rangle^2\langle\frac{\alpha_sGG}{\pi}\rangle$. Other diagrams obtained by interchanging of the light quark lines and  heavy quark lines are implied.}\label{nonfac-qqg-qqg}
\end{figure}

In the QCD sum rules, we carry out the operator product expansion at the deep Euclidean region $P^2=-p^2\to\infty$ or $\gg \Lambda^2_{QCD}$, which corresponds to $x_0\sim \vec{x}\sim \frac{1}{\sqrt{P^2}}$, and $x^2\sim \frac{1}{P^2}$, it is questionable to apply the Landau equation to study the Feynman diagrams \cite{Landau}.   At the QCD side, the correlation function $\Pi_{\mu\nu}(p)$ of the order $\mathcal{O}(\epsilon^0)$ can be written as
\begin{eqnarray}\label{CF-QCD-side}
\Pi_{\mu\nu}(p)&=&-\frac{i\varepsilon^{ijk}\varepsilon^{imn}\varepsilon^{i^{\prime}j^{\prime}k^{\prime}}\varepsilon^{i^{\prime}m^{\prime}n^{\prime}}}{2}\int d^4x e^{ip \cdot x}   \nonumber\\
&&\left\{{\rm Tr}\left[ \gamma_5C^{kk^{\prime}}(x)\gamma_5 CU^{jj^{\prime}T}(x)C\right] {\rm Tr}\left[ \gamma_\nu C^{n^{\prime}n}(-x)\gamma_\mu C D^{m^{\prime}mT}(-x)C\right] \right. \nonumber\\
&&+{\rm Tr}\left[ \gamma_\mu C^{kk^{\prime}}(x)\gamma_\nu CU^{jj^{\prime}T}(x)C\right] {\rm Tr}\left[ \gamma_5 C^{n^{\prime}n}(-x)\gamma_5 C D^{m^{\prime}mT}(-x)C\right] \nonumber\\
&&+{\rm Tr}\left[ \gamma_\mu C^{kk^{\prime}}(x)\gamma_5 CU^{jj^{\prime}T}(x)C\right] {\rm Tr}\left[ \gamma_\nu C^{n^{\prime}n}(-x)\gamma_5 C D^{m^{\prime}mT}(-x)C\right] \nonumber\\
 &&\left.+{\rm Tr}\left[ \gamma_5 C^{kk^{\prime}}(x)\gamma_\nu CU^{jj^{\prime}T}(x)C\right] {\rm Tr}\left[ \gamma_5 C^{n^{\prime}n}(-x)\gamma_\mu C D^{m^{\prime}mT}(-x)C\right] \right\} \, ,
\end{eqnarray}
where
the $U_{ij}(x)$, $D_{ij}(x)$ and $C_{ij}(x)$ are the full $u$, $d$ and $c$ quark propagators respectively ($S_{ij}(x)=U_{ij}(x),\,D_{ij}(x)$),
 \begin{eqnarray}
S_{ij}(x)&=& \frac{i\delta_{ij}\!\not\!{x}}{ 2\pi^2x^4}-\frac{\delta_{ij}\langle
\bar{q}q\rangle}{12} -\frac{\delta_{ij}x^2\langle \bar{q}g_s\sigma Gq\rangle}{192} -\frac{ig_sG^{a}_{\alpha\beta}t^a_{ij}(\!\not\!{x}
\sigma^{\alpha\beta}+\sigma^{\alpha\beta} \!\not\!{x})}{32\pi^2x^2} \nonumber\\
&&  -\frac{\delta_{ij}x^4\langle \bar{q}q \rangle\langle g_s^2 GG\rangle}{27648} -\frac{1}{8}\langle\bar{q}_j\sigma^{\mu\nu}q_i \rangle \sigma_{\mu\nu}+\cdots \, ,
\end{eqnarray}
\begin{eqnarray}
C_{ij}(x)&=&\frac{i}{(2\pi)^4}\int d^4k e^{-ik \cdot x} \left\{
\frac{\delta_{ij}}{\!\not\!{k}-m_c}
-\frac{g_sG^n_{\alpha\beta}t^n_{ij}}{4}\frac{\sigma^{\alpha\beta}(\!\not\!{k}+m_c)+(\!\not\!{k}+m_c)
\sigma^{\alpha\beta}}{(k^2-m_c^2)^2}\right.\nonumber\\
&&\left. -\frac{g_s^2 (t^at^b)_{ij} G^a_{\alpha\beta}G^b_{\mu\nu}(f^{\alpha\beta\mu\nu}+f^{\alpha\mu\beta\nu}+f^{\alpha\mu\nu\beta}) }{4(k^2-m_c^2)^5}+\cdots\right\} \, ,\nonumber\\
f^{\alpha\beta\mu\nu}&=&(\!\not\!{k}+m_c)\gamma^\alpha(\!\not\!{k}+m_c)\gamma^\beta(\!\not\!{k}+m_c)\gamma^\mu(\!\not\!{k}+m_c)\gamma^\nu(\!\not\!{k}+m_c)\, ,
\end{eqnarray}
and  $t^n=\frac{\lambda^n}{2}$, the $\lambda^n$ is the Gell-Mann matrix   \cite{WangHuangTao-3900,PRT85,Pascual-1984}.

As there exists a spatial distance $\epsilon$ between the diquark and antidiquark constituents,  we split the point $0$ (and $x$) into two points,
\begin{eqnarray}
0&\to & 0, \,\, 0+\epsilon\, , \nonumber\\
x&\to & x, \,\, x+\epsilon\, ,
\end{eqnarray}
to distinguish the  diquark and antidiquark contributions in drawing the Feynman diagrams according to Eq.\eqref{CF-QCD-side}.
We classify the Feynman diagrams as factorizable diagrams and nonfactorizable  diagrams respectively.
In Figs.\ref{Lowest-diagram}-\ref{fac-qqg-qqg}, we draw the factorizable Feynman diagrams, in which the contributions come from the diquark loops and antidiquark loops are
factorizable due to the nonzero spatial distance $\epsilon$. They contribute to the perturbative terms, $\langle\bar{q}q\rangle$,  $\langle\bar{q}g_{s}\sigma Gq\rangle$, $\langle\frac{\alpha_{s}GG}{\pi}\rangle$,  $\langle\bar{q}q\rangle^2$,
$\langle\bar{q}q\rangle \langle\frac{\alpha_{s}GG}{\pi}\rangle$,  $\langle\bar{q}q\rangle  \langle\bar{q}g_{s}\sigma Gq\rangle$,
$\langle\bar{q}q\rangle^2 \langle\frac{\alpha_{s}GG}{\pi}\rangle$ and $\langle\bar{q}g_{s}\sigma Gq\rangle^2$.
In Figs.\ref{nonfac-qqg}-\ref{nonfac-qqg-qqg}, we draw the nonfactorizable Feynman diagrams, in which the contributions come from the diquark loops and antidiquark loops are nonfactorizable even if we take into account  the nonzero spatial distance $\epsilon$. They contribute to the $\langle\bar{q}g_{s}\sigma Gq\rangle$, $\langle\frac{\alpha_{s}GG}{\pi}\rangle$,
$\langle\bar{q}q\rangle \langle\frac{\alpha_{s}GG}{\pi}\rangle$,  $\langle\bar{q}q\rangle  \langle\bar{q}g_{s}\sigma Gq\rangle$,
$\langle\bar{q}q\rangle^2 \langle\frac{\alpha_{s}GG}{\pi}\rangle$ and $\langle\bar{q}g_{s}\sigma Gq\rangle^2$.

We compute both the factorizable  and nonfactorizable Feynman diagrams, and obtain the correlation function $\Pi(p^2)$ at the quark level, then obtain the QCD spectral density through dispersion relation. We match the hadron side with  the QCD side of the correlation function $\Pi(p^2)$  below the continuum threshold parameter $s_0$, then perform the Borel transformation with respect to the variable $P^2=-p^2$ to obtain the QCD sum rules,
\begin{eqnarray}\label{QCD-fac-plus-nonfac}
\lambda_Z^2\exp\left(-\frac{M_Z^2}{T^2} \right)+\Pi_{\rm RC}(T^2)&=& \int_{4m_c^2}^{s_0}ds\,\Big[\rho_{f}(s)+\rho_{nf}(s)\Big]\exp\left(-\frac{s}{T^2} \right)\, ,
\end{eqnarray}
where the two particle scattering state  contributions (RC),
\begin{eqnarray}
\Pi_{\rm RC}(T^2)&=&\lambda_{\eta_c \rho;11}^2\int_{m^2_{\eta_c \rho}}^{s_0}ds \frac{\sqrt{\lambda(s,m_{\eta_c}^2,m^2_{\rho})}}{s}\left[1+\frac{\lambda(s,m_{\eta_c}^2,m^2_{\rho})}{12sm_\rho^2} \right]\exp\left(-\frac{s}{T^2}\right)\nonumber\\
&&+\lambda_{\pi J/\psi;22}^2\int_{m^2_{\pi J/\psi}}^{s_0}ds \frac{\sqrt{\lambda(s,m_{\pi}^2,m^2_{J/\psi})}}{s}\left[1+\frac{\lambda(s,m_{\pi}^2,m^2_{J/\psi})}{12sm_{J/\psi}^2} \right]\exp\left(-\frac{s}{T^2}\right)\nonumber\\
&&+\frac{\lambda_{\eta_c \rho;33}^2}{4}\int_{m^2_{\eta_c \rho}}^{s_0}ds \frac{\sqrt{\lambda(s,m_{\eta_c}^2,m^2_{\rho})}}{s}\left[(s-m_{\eta_c}^2-m^2_{\rho})^2-\frac{\lambda(s,m_{\eta_c}^2,m^2_{\rho})(s-m_\rho^2)}{3s} \right]\exp\left(-\frac{s}{T^2}\right)\nonumber\\
&&+\frac{\lambda_{\pi J/\psi;44}^2}{4}\int_{m^2_{\pi J/\psi}}^{s_0}ds \frac{\sqrt{\lambda(s,m_{\pi}^2,m^2_{J/\psi})}}{s}\left[(s-m_{\pi}^2-m^2_{J/\psi})^2-\frac{\lambda(s,m_{\pi}^2,m^2_{J/\psi})(s-m_{J/\psi}^2)}{3s} \right]\exp\left(-\frac{s}{T^2}\right)\nonumber\\
&&+\lambda_{\eta_c \rho;13}^2\int_{m^2_{\eta_c \rho}}^{s_0}ds \frac{\sqrt{\lambda(s,m_{\eta_c}^2,m^2_{\rho})}}{s}\left[\frac{\lambda(s,m_{\eta_c}^2,m^2_{\rho})}{6}-(s-m_{\eta_c}^2-m^2_{\rho}) \right]\exp\left(-\frac{s}{T^2}\right)\nonumber\\
&&+\lambda_{\pi J/\psi;24}^2\int_{m^2_{\pi J/\psi}}^{s_0}ds \frac{\sqrt{\lambda(s,m_{\pi}^2,m^2_{J/\psi})}}{s}\left[\frac{\lambda(s,m_{\pi}^2,m^2_{J/\psi})}{6}-(s-m_{\pi}^2-m^2_{J/\psi}) \right]\exp\left(-\frac{s}{T^2}\right)\nonumber\\
&&+\lambda_{\eta_c b_1;33}^2\int_{m^2_{\eta_c b_1}}^{s_0}ds \frac{\sqrt{\lambda(s,m_{\eta_c}^2,m^2_{b_1})}}{s}\frac{\lambda(s,m_{\eta_c}^2,m^2_{b_1})}{6} \exp\left(-\frac{s}{T^2}\right)\nonumber
\end{eqnarray}
\begin{eqnarray}
&&+\lambda_{\chi_{c1} \rho;33}^2\int_{m^2_{\chi_{c1}  \rho}}^{s_0}ds \frac{\sqrt{\lambda(s,m_{\chi_{c1} }^2,m^2_{\rho})}}{s}\frac{\lambda(s,m_{\chi_{c1}}^2,m^2_{\rho})(2s+m_\rho^2+2m^2_{\chi_{c1}})}{12sm^2_{\chi_{c1}}}\exp\left(-\frac{s}{T^2}\right)\nonumber\\
&&+\lambda_{\pi h_c;44}^2\int_{m^2_{\pi h_c}}^{s_0}ds \frac{\sqrt{\lambda(s,m_{\pi}^2,m^2_{h_c})}}{s}\frac{\lambda(s,m_{\pi}^2,m^2_{h_c})}{6} \exp\left(-\frac{s}{T^2}\right)\nonumber\\
&&+\lambda_{a_{1} J/\psi;44}^2\int_{m^2_{a_{1}  J/\psi}}^{s_0}ds \frac{\sqrt{\lambda(s,m_{a_{1} }^2,m^2_{J/\psi})}}{s}\frac{\lambda(s,m_{a_{1}}^2,m^2_{J/\psi})(2s+m_{J/\psi}^2+2m^2_{a_{1}})}{12sm^2_{a_{1}}}\exp\left(-\frac{s}{T^2}\right)\nonumber\\
&&+\lambda_{D_0 D;55/66}^2\int_{m^2_{D_0D}}^{s_0}ds \frac{\sqrt{\lambda(s,m_{D_0}^2,m^2_{D})}}{s}\frac{\lambda(s,m_{D_0}^2,m^2_{D})}{12s}\exp\left(-\frac{s}{T^2}\right)\nonumber\\
&&+\lambda_{D^*D^*;77/88}^2\int_{m^2_{D^*D^*}}^{s_0}ds \frac{\sqrt{\lambda(s,m_{D^*}^2,m^2_{D^*})}}{s}\left[2m_{D^*}^2+\frac{\lambda(s,m_{D^*}^2,m^2_{D^*})(s+m_{D^*}^2)}{6sm_{D^*}^2} \right]\exp\left(-\frac{s}{T^2}\right)\nonumber\\
&&+\lambda_{D_0D^*;77/88}^2\int_{m^2_{D_0D^*}}^{s_0}ds \frac{\sqrt{\lambda(s,m_{D_0}^2,m^2_{D^*})}}{s}\frac{\lambda(s,m_{D_0}^2,m^2_{D^*})}{6}\exp\left(-\frac{s}{T^2} \right)\nonumber\\
&&+\lambda_{\pi \psi^\prime;22}^2\int_{m^2_{\pi \psi^\prime}}^{s_0}ds \frac{\sqrt{\lambda(s,m_{\pi}^2,m^2_{\psi^\prime})}}{s}\left[1+\frac{\lambda(s,m_{\pi}^2,m^2_{\psi^\prime})}{12sm_{\psi^\prime}^2} \right]\exp\left(-\frac{s}{T^2} \right)\nonumber\\
&&+\frac{\lambda_{\pi \psi^\prime;44}^2}{4}\int_{m^2_{\pi \psi^\prime}}^{s_0}ds \frac{\sqrt{\lambda(s,m_{\pi}^2,m^2_{\psi^\prime})}}{s}\left[(s-m_{\pi}^2-m^2_{\psi^\prime})^2-\frac{\lambda(s,m_{\pi}^2,m^2_{\psi^\prime})(s-m_{\psi^\prime}^2)}{3s} \right]\exp\left(-\frac{s}{T^2} \right)\nonumber\\
&&+\lambda_{\pi \psi^\prime;24}^2\int_{m^2_{\pi \psi^\prime}}^{s_0}ds \frac{\sqrt{\lambda(s,m_{\pi}^2,m^2_{\psi^\prime})}}{s}\left[\frac{\lambda(s,m_{\pi}^2,m^2_{\psi^\prime})}{6}-(s-m_{\pi}^2-m^2_{\psi^\prime}) \right]\exp\left(-\frac{s}{T^2} \right)\nonumber\\
&&+\lambda_{\pi h_c;44}^2\int_{m^2_{\pi h^\prime_c}}^{s_0}ds \frac{\sqrt{\lambda(s,m_{\pi}^2,m^2_{h^\prime_c})}}{s}\frac{\lambda(s,m_{\pi}^2,m^2_{h^\prime_c})}{6}\exp\left(-\frac{s}{T^2} \right)+\left(\pi \psi^\prime \to \pi \psi^{\prime\prime}  \right)\, ,
\end{eqnarray}
the QCD spectral densities $\rho_{f}(s)$ and $\rho_{nf}(s)$ receive contributions from the factorizable  and nonfactorizable Feynman diagrams, respectively. The explicit expressions of the QCD spectral densities $\rho_{f}(s)$ and $\rho_{nf}(s)$ are available upon request by contacting me via  E-mail.

If there exists a repulsive barrier or spatial distance between the diquark and antidiquark constituents, the Feynman diagrams can be divided into factorizable  and nonfactorizable diagrams. The factorizable Feynman diagrams correspond to the stable diquark-antidiquark type contributions.
The repulsive barrier or spatial distance frustrates  dissociation of the diquark and antidiquark states to form the color singlet quark-antiquark pairs $q\bar{q}$, $Q\bar{Q}$ or $Q\bar{q}$, $q\bar{Q}$, the colored diquark and antidiquark constituents are confined objects, which lead to
a stable tetraquark state.
The nonfactorizable Feynman diagrams correspond to the tunnelling effects between diquark and antidiquark constituents, and facilitate dissociation of the diquark and antidiquark states to form the color singlet quark-antiquark pairs $q\bar{q}$, $Q\bar{Q}$ or $Q\bar{q}$, $q\bar{Q}$. In this case, the QCD sum rules in Eq.\eqref{QCD-fac-plus-nonfac} can be replaced with two QCD sum rules,
\begin{eqnarray}\label{QCD-fac-nonfac-Zc}
\lambda_Z^2\exp\left(-\frac{M_Z^2}{T^2} \right)&=& \int_{4m_c^2}^{s_0}ds\,\rho_{f}(s)\,\exp\left(-\frac{s}{T^2} \right)\, ,
\end{eqnarray}

\begin{eqnarray}\label{QCD-fac-nonfac-two-par}
\Pi_{\rm RC}(T^2)&=& \int_{4m_c^2}^{s_0}ds\,\rho_{nf}(s)\, \exp\left(-\frac{s}{T^2} \right)\, .
\end{eqnarray}

On the other hand, if the non-local effects of the repulsive barrier or spatial distance between the diquark and antidiquark constituents  are neglectful,
 we can set the finite four-vector $\epsilon^\alpha=0$ and perform Fierz rearrangement for the diquark-antidiquark type current in the color and Dirac-spinor  spaces freely to obtain a special superposition of the     color-singlet-color-singlet type or meson-meson type currents, which couple potentially to the meson-meson pairs, it is not necessary to take into account the pole term of the $Z_c(3900)$, in other words, the axialvector tetraquark state doest not exist as a real resonance, it is a virtual state and embodies the net effects. And it is not necessary to divide
 the Feynman diagrams  into factorizable  and nonfactorizable parts. We saturate the hadron side of the QCD sum rules with the two-particle scattering state contributions,
  \begin{eqnarray}\label{QCD-Two-Particle}
\Pi_{\rm RC}(T^2)&=& \kappa \int_{4m_c^2}^{s_0}ds\,\Big[\rho_{f}(s)+\rho_{nf}(s)\Big]\exp\left(-\frac{s}{T^2} \right)\, ,
\end{eqnarray}
where we introduce a parameter $\kappa$ to measure  the deviation from the ideal value $1$.

We derive Eq.\eqref{QCD-fac-plus-nonfac}, Eq.\eqref{QCD-fac-nonfac-Zc} and Eq.\eqref{QCD-Two-Particle} with respect to $\tau=\frac{1}{T^2}$ , and obtain the QCD sum rules for the tetraquark mass and other QCD sum rules,
\begin{eqnarray}
M_Z^2&=& \frac{-\frac{d}{d\tau}\Big\{\int_{4m_c^2}^{s_0}ds\,\Big[\rho_{f}(s)+\rho_{nf}(s)\Big]\exp\left(-\frac{s}{T^2} \right)-\Pi_{\rm RC}(T^2)\Big\}}{\int_{4m_c^2}^{s_0}ds\,\Big[\rho_{f}(s)+\rho_{nf}(s)\Big]\exp\left(-\frac{s}{T^2} \right)-\Pi_{\rm RC}(T^2)}\mid_{\tau=\frac{1}{T^2}}\, ,
\end{eqnarray}

\begin{eqnarray}\label{QCD-fac-nonfac-Zc-derive}
M_Z^2&=& \frac{-\frac{d}{d\tau}\int_{4m_c^2}^{s_0}ds\,\rho_{f}(s)\exp\left(-\frac{s}{T^2} \right)}{\int_{4m_c^2}^{s_0}ds\,\rho_{f}(s)\exp\left(-\frac{s}{T^2} \right)}\mid_{\tau=\frac{1}{T^2}}\, ,
\end{eqnarray}

 \begin{eqnarray}\label{QCD-Two-Particle-derive}
\frac{d}{d\tau}\Pi_{\rm RC}(T^2)&=& \kappa \frac{d}{d\tau}\int_{4m_c^2}^{s_0}ds\,\Big[\rho_{f}(s)+\rho_{nf}(s)\Big]\exp\left(-\frac{s}{T^2} \right)\mid_{\tau=\frac{1}{T^2}}\, .
\end{eqnarray}

 For the color-singlet-color-singlet type or  meson-meson type  currents,  Lucha, Melikhov and Sazdjian assert that the contributions  at the order $\mathcal{O}(\alpha_s^k)$ with $k\leq1$ in the operator product expansion, which are factorizable in the color space, are exactly  canceled out   by the meson-meson scattering states at the hadron side, the tetraquark molecular states begin to receive contributions at the order $\mathcal{O}(\alpha_s^2)$  \cite{Chu-Sheng-PRD,Chu-Sheng-EPJC}. If the assertion  of  Lucha, Melikhov and Sazdjian is  right, only the two QCD sum rules shown in Eq.\eqref{QCD-Two-Particle} and Eq.\eqref{QCD-Two-Particle-derive} survive in the present case according to the Fierz rearrangement in Eq.\eqref{Fierz}.

\begin{figure}
 \centering
  \includegraphics[totalheight=4cm,width=5cm]{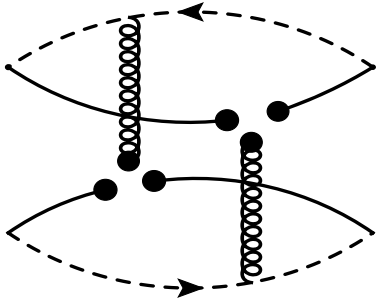}
 \caption{ The connected   Feynman diagrams contribute to the vacuum condensates $\langle \bar{q}g_s\sigma G q \rangle^2$ for the meson-meson type currents. }\label{meson-qqg-qqg}
\end{figure}

The vacuum condensates serve as a  landmark for  the nonperturbative nature of the QCD sum rules,  the connected (or nonfactorizable) contributions in the case of meson-meson type  currents appear at the  order $\mathcal{O}(\alpha_s)$ due to the   operators  $\bar{q}g_sGq\bar{q}g_sGq$, which  come from the Feynman diagrams shown in Fig.\ref{meson-qqg-qqg} and lead to the vacuum condensate $\langle\bar{q}g_s\sigma Gq\rangle^2$,  those contributions  are of the order $\mathcal{O}(\alpha_s^0)$ as a matter of fact. The tetraquark molecular states begin to receive connected (or nonfactorizable) contributions at the order   $\mathcal{O}(\alpha_s^0)$, not at the order $\mathcal{O}(\alpha_s^2)$ asserted in Refs.\cite{Chu-Sheng-PRD,Chu-Sheng-EPJC}.

 Lucha, Melikhov and Sazdjian use the Landau equation to study the Feynman diagrams in the QCD sum rules \cite{Chu-Sheng-PRD,Chu-Sheng-EPJC}.  In fact,  the quarks and gluons are confined objects, they cannot be put on the mass-shell, it is questionable  to assert  that the Landau equation is applicable  in the nonperturbative  calculations exploring  the four-quark bound states \cite{Landau}. The Landau singularity  is just a signal of a four-quark intermediate state, not a signal of  a tetraquark (molecular) state or two-meson scattering state, because  the Landau singularity  is merely  a  kinematical singularity, not a dynamical singularity \cite{FKGuo-cusps}.

We should bear in mind that the Landau equation requires   pole masses rather than  $\overline{MS}$ masses to ensure  that  there  exist  mass poles  or mass-shells in pure perturbative calculations. For the heavy quarks, the pole masses are  $\hat{m}_c=1.67\pm0.07\,\rm{GeV}$ and $\hat{m}_b=4.78\pm0.06\,\rm{GeV}$ from the Particle Data Group \cite{PDG}, the thresholds $2\hat{m}_c=3.34\pm0.14\,{\rm{GeV}}>m_{\eta_c}$,  $m_{J/\psi}$, and $2\hat{m}_b=9.56\pm0.12\,{\rm{GeV}}>m_{\eta_b}$, $m_{\Upsilon}$.
According to the assertion of Lucha, Melikhov and Sazdjian, the tetraquark (molecular) states begin to receive contributions at the order $\mathcal{O}(\alpha_s^2)$,
it is reasonable to choose the values of the pole masses from the Particle Data Group, which receive contributions from the divergent power series in perturbative  expansions up to the order $\mathcal{O}(\alpha_s^3)$.
It is  unreliable  that the  masses of the ground state charmonium (bottomonium) states lie below the threshold $2\hat{m}_c$ ($2\hat{m}_b$) in the QCD sum rules for the $\eta_c$ and $J/\psi$ ($\eta_b$ and $\Upsilon$).

In Refs.\cite{Chu-Sheng-PRD,Chu-Sheng-EPJC}, Lucha, Melikhov and Sazdjian only obtain formal QCD sum rules for the tetraquark (molecular) states, do  not obtain
 feasible QCD sum rules with predictions can be confronted  to the experimental data.
For the meson-meson type currents, it is reasonable to take the viewpoint that the disconnected Feynman diagrams give masses to the meson-meson pair,   while the connected Feynman diagrams contribute weak  attractive interactions to bind the massive meson-meson pair  to form loosely molecular states.

\section{Numerical results and discussions}
At the QCD side, we take  the standard values of the vacuum condensates $\langle
\bar{q}q \rangle=-(0.24\pm 0.01\, \rm{GeV})^3$,   $\langle
\bar{q}g_s\sigma G q \rangle=m_0^2\langle \bar{q}q \rangle$,
$m_0^2=(0.8 \pm 0.1)\,\rm{GeV}^2$,   $\langle \frac{\alpha_s
GG}{\pi}\rangle=(0.33\,\rm{GeV})^4 $    at the energy scale  $\mu=1\, \rm{GeV}$
\cite{PRT85,SVZ79,ColangeloReview}, and choose the $\overline{MS}$ mass  $m_{c}(m_c)=(1.275\pm0.025)\,\rm{GeV}$
 from the Particle Data Group \cite{PDG}.
Moreover, we take into account the energy-scale dependence of  the  parameters,
\begin{eqnarray}
\langle\bar{q}q \rangle(\mu)&=&\langle\bar{q}q \rangle({\rm 1 GeV})\left[\frac{\alpha_{s}({\rm 1 GeV})}{\alpha_{s}(\mu)}\right]^{\frac{12}{25}}\, , \nonumber\\
 \langle\bar{q}g_s \sigma Gq \rangle(\mu)&=&\langle\bar{q}g_s \sigma Gq \rangle({\rm 1 GeV})\left[\frac{\alpha_{s}({\rm 1 GeV})}{\alpha_{s}(\mu)}\right]^{\frac{2}{25}}\, , \nonumber\\
m_c(\mu)&=&m_c(m_c)\left[\frac{\alpha_{s}(\mu)}{\alpha_{s}(m_c)}\right]^{\frac{12}{25}} \, ,\nonumber\\
\alpha_s(\mu)&=&\frac{1}{b_0t}\left[1-\frac{b_1}{b_0^2}\frac{\log t}{t} +\frac{b_1^2(\log^2{t}-\log{t}-1)+b_0b_2}{b_0^4t^2}\right]\, ,
\end{eqnarray}
   where $t=\log \frac{\mu^2}{\Lambda^2}$, $b_0=\frac{33-2n_f}{12\pi}$, $b_1=\frac{153-19n_f}{24\pi^2}$,
   $b_2=\frac{2857-\frac{5033}{9}n_f+\frac{325}{27}n_f^2}{128\pi^3}$,
   $\Lambda=210\,\rm{MeV}$, $292\,\rm{MeV}$  and  $332\,\rm{MeV}$ for the flavors
   $n_f=5$, $4$ and $3$, respectively  \cite{PDG,Narison-mix}, and evolve all the parameters to the ideal   energy scale   $\mu$  with $n_f=4$ to extract the
 axialvector   tetraquark mass as the $c$-quark is concerned.

At the hadron side, we take the hadronic parameters  as
$m_{\eta_c}=2.9839\,\rm{GeV}$,
$m_{J/\psi}=3.0969\,\rm{GeV}$,
$m_{h_c}=3.5254\,\rm{GeV}$,
$m_{\chi_{c1}}=3.5107\,\rm{GeV}$,
$m_{\rho}=0.7753\,\rm{GeV}$,
$m_{a_1}=1.2300\,\rm{GeV}$,
$m_{b_1}=1.2295\,\rm{GeV}$,
$m_{\pi}=0.1396\,\rm{GeV}$,
$m_{D}=1.8672\,\rm{GeV}$,
$m_{D^*}=2.0086\,\rm{GeV}$,
$m_{D_0}=2.3245\,\rm{GeV}$,
$m_{\psi^\prime}=3.6861\,\rm{GeV}$,
$m_{\psi^{\prime\prime}}=4.0396\,\rm{GeV}$ from the Particle Data Group \cite{PDG};
$m_{h^\prime_c}=3.9560\,\rm{GeV}$ for the Godfrey-Isgur model  \cite{Godfrey-hc},
$f_{J/\psi}=0.418 \,\rm{GeV}$,
$f_{\eta_c}=0.387 \,\rm{GeV}$,
$f_{J/\psi}^T=0.410 \,\rm{GeV}$,
$f_{h_c}=0.235 \,\rm{GeV}$
 from Lattice QCD \cite{Becirevic};
 $f_{\chi_{c1}}=0.338 \,\rm{GeV}$ \cite{VANovikov-PRT},
 $f_{\rho}=0.205 \,\rm{GeV}$,
 $f_{\rho}^T=0.160 \,\rm{GeV}$ \cite{PBall2004},
  $f_{b_1}=0.180 \,\rm{GeV}$ \cite{PBall1996},
   $f_{a_1}=0.238 \,\rm{GeV}$ \cite{KCYang},
 $f_{D}=0.208 \,\rm{GeV}$,
$f_{D^*}=0.263 \,\rm{GeV}$,
$f_{D_0}=0.373 \,\rm{GeV}$ \cite{Wang-DecayConst} from the QCD sum rules;
 $f_{\pi}=0.130 \,\rm{GeV}$, $f_{\psi^\prime}=0.295 \,\rm{GeV}$, $f_{\psi^{\prime\prime}}=0.187 \,\rm{GeV}$ extracted from the experimental data \cite{PDG};
 $f_{D^*}^T=f_{D^*}$, $f_{\psi^\prime}^T=f_{\psi^\prime}$, $f^T_{\psi^{\prime\prime}}=f_{\psi^{\prime\prime}}$,
  $f_{h_c^\prime}=f_{h_c}\frac{f_{\psi^\prime}}{f_{J/\psi}}=0.166\,\rm{GeV}$ estimated in the present work;
$f_{\pi}m^2_{\pi}/(m_u+m_d)=-2\langle \bar{q}q\rangle/f_{\pi}$ from the Gell-Mann-Oakes-Renner relation. In the present work, we take the average values of the charged and neutral $D$ mesons.

We choose the values of the decay constants from the lattice QCD, the QCD sum rules or extracted from the experimental data, where the perturbative corrections have been taken into account. In the present work, we obtain the expressions of the QCD spectral densities $\rho_f(s)$ and $\rho_{nf}(s)$ at  the leading order, the value of the
$\Pi_{\rm RC}(T^2)$ at the hadron side is overestimated. At the leading order, $\overline{f}_{D}=0.170\,\rm{GeV}$, $\overline{f}_{D^*}=0.240\,\rm{GeV}$ \cite{Khodjamirian-1998}.
The $\Pi_{\rm RC}(T^2)\propto f_M^4$, where the $M$ denotes the mesons. At the leading order, we can take the replacement  $\Pi_{\rm RC}(T^2)\to \overline{\Pi}_{\rm RC}(T^2)=\Pi_{\rm RC}(T^2) \frac{\overline{f}^4_M}{f^4_M}$. Considering the ratios  $\frac{\overline{f}^4_D}{f^4_D}=0.45$ and $\frac{\overline{f}^4_{D^*}}{f^4_{D^*}}=0.69$, we multiply the $\Pi_{\rm RC}(T^2)$ by a factor $0.7$ to subtract the perturbative corrections  mildly, $\overline{\Pi}_{\rm RC}(T^2)=0.7\,\Pi_{\rm RC}(T^2)$.

\subsection{$Z_c(3900)$ plus two-particle scattering states}

From Table \ref{meson-meson-pair}, we can see that the largest thresholds are $m_{a_1J/\psi}$ and $m_{D_0D^*}$. We take into account the large widths of the scalar $D$ mesons, $\Gamma_{D_0^0}=274\pm40\,\rm{MeV}$ and $\Gamma_{D_0^\pm}=221\pm18\,\rm{MeV}$ \cite{PDG}, and choose the continuum threshold parameter as $\sqrt{s_0}=M_{Z_c}+0.55\,\rm{GeV}=4.45\,\rm{GeV}$ according to mass gaps $M_{Z_c(4430)}-M_{Z_c(3900)}=m_{\psi^\prime}-m_{J/\psi} ={\rm 589\,MeV}$ \cite{PDG}.
As the LHCb collaboration confirmed  the $Z^-_c(4430)$ state in the $B^0\to\psi'\pi^-K^+$ decays
and established its spin-parity  $J^P=1^+$  \cite{LHCb-1404}, we can assign the $Z_c(3900)$ and $Z_c(4430)$  to be the ground state and the first radial excitation  of the axialvector tetraquark states  respectively  according to the analogous decays, $Z_c(3900)^\pm \to J/\psi\pi^\pm$, $Z_c(4430)^\pm \to \psi^\prime\pi^\pm$,
and the mass gaps $M_{Z_c(4430)}-M_{Z_c(3900)}= m_{\psi^\prime}-m_{J/\psi}$ \cite{Z4430-1405,Nielsen-1401,Wang4430,Azizi-4430}.
Furthermore, we add  a uncertainty $0.1\,\rm{GeV}$. If only the $Z_c(3900)$ is retained, we can choose a slightly smaller continuum threshold parameter $\sqrt{s_0}=M_{Z_c}+0.50\pm0.1\,\rm{GeV}=4.40\pm0.1\,\rm{GeV}$.
At the optimal energy scale $\mu=1.4\,\rm{GeV}$ \cite{Wang-Hidden-charm}, the pole contribution (PC) is,
\begin{eqnarray}
{\rm PC}&=&\frac{\int_{4m_c^2}^{s_0}ds\,\Big[\rho_{f}(s)+\rho_{nf}(s)\Big]\exp\left(-\frac{s}{T^2} \right)}{\int_{4m_c^2}^{\infty}ds\,\Big[\rho_{f}(s)+\rho_{nf}(s)\Big]\exp\left(-\frac{s}{T^2} \right)} \nonumber\\
&=&(43-66)\% \, ,
\end{eqnarray}
at the Borel window $T^2=(2.7-3.1)\,\rm{GeV}$. The pole dominance criterion is well satisfied. On the other hand, the contributions of the vacuum condensates of dimension $10$ are less than $1\%$, the operator product expansion is well convergent.

\begin{figure}
 \centering
 \includegraphics[totalheight=6cm,width=7cm]{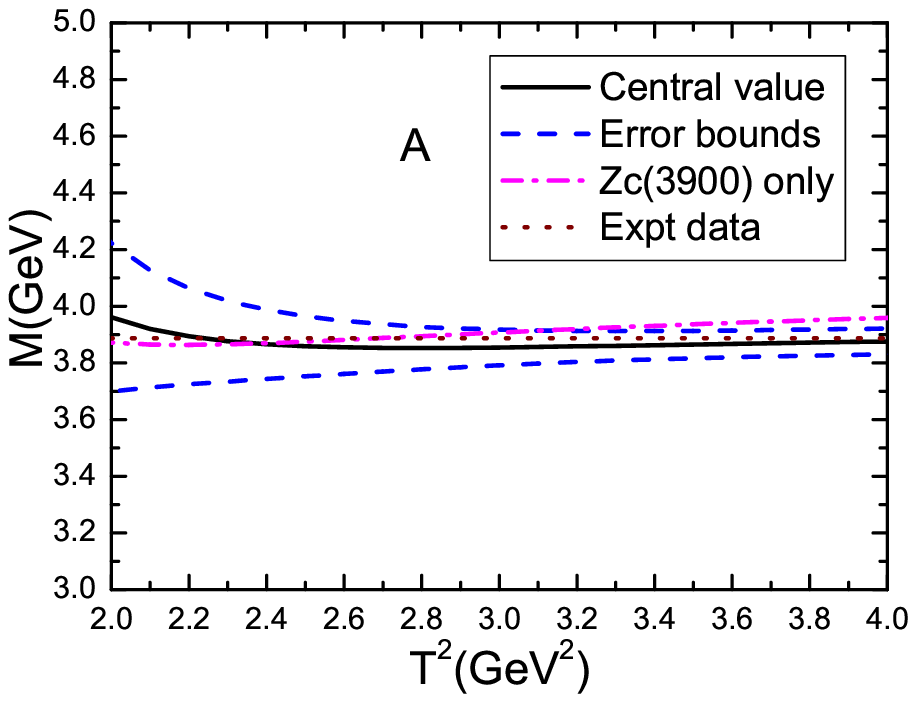}
 \includegraphics[totalheight=6cm,width=7cm]{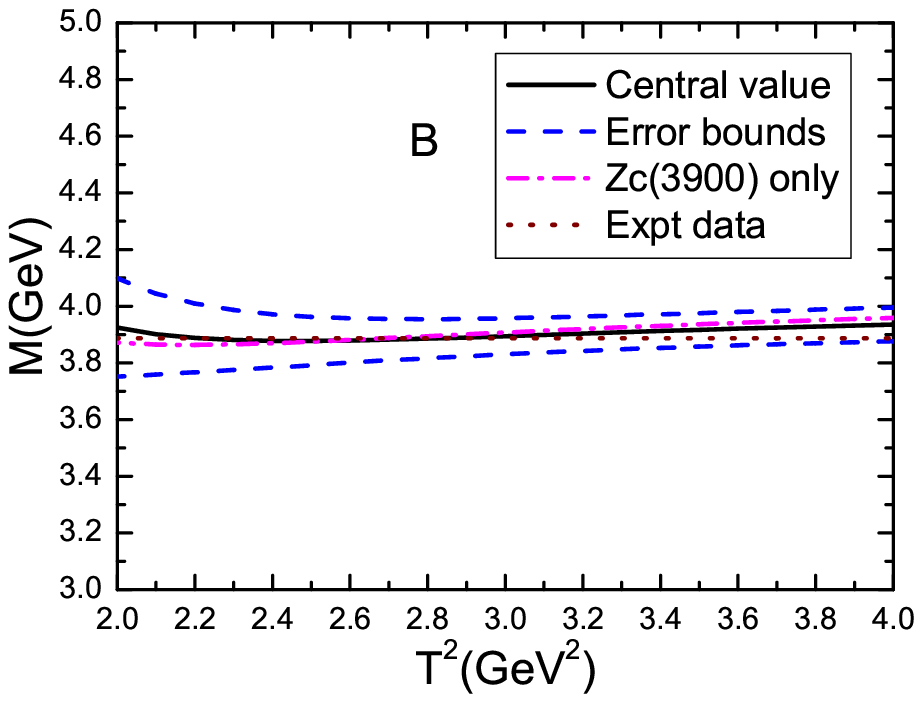}
 \caption{ The mass of the  $Z_c(3900)$ with variations  of the Borel parameter $T^2$, where the $A$ and $B$ correspond to the $\Pi_{\rm RC}(T^2)$ and $\overline{\Pi}_{\rm RC}(T^2)$, respectively.  }\label{mass-Zc-MM}
\end{figure}
\begin{figure}
 \centering
 \includegraphics[totalheight=6cm,width=7cm]{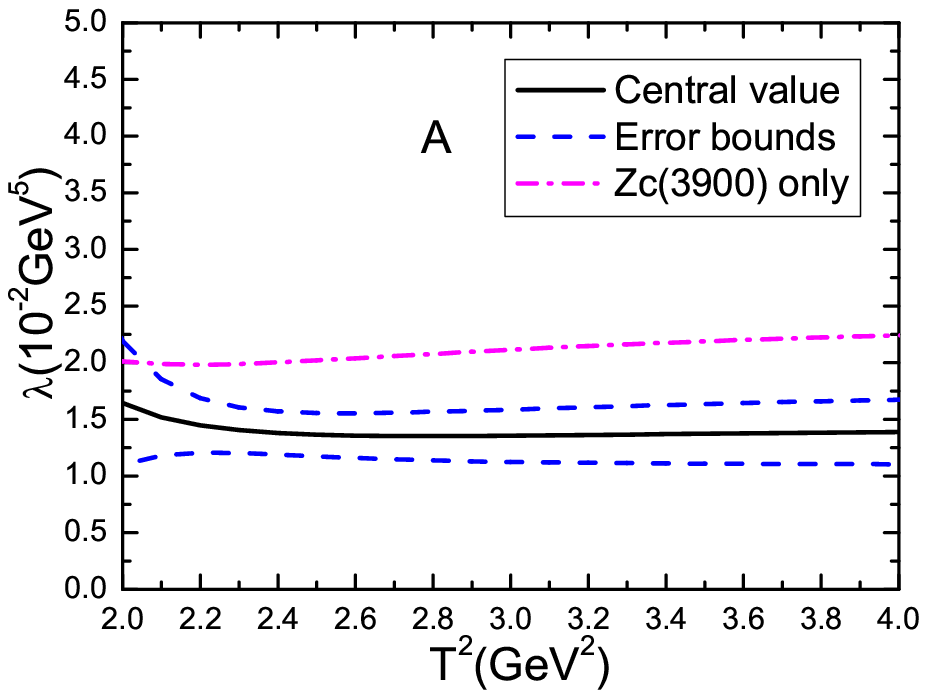}
 \includegraphics[totalheight=6cm,width=7cm]{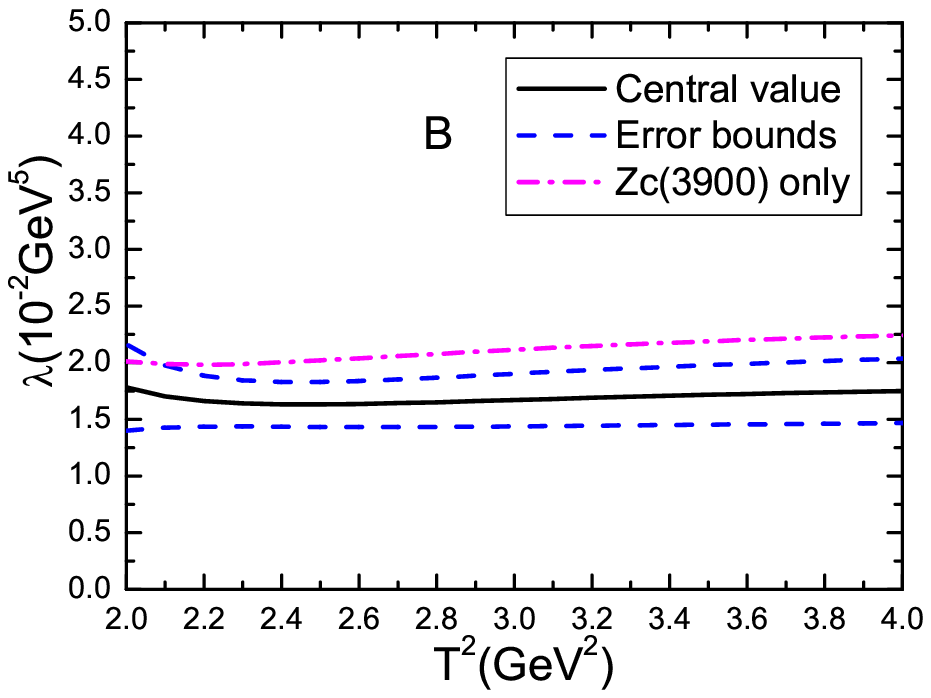}
 \caption{ The pole residue of the  $Z_c(3900)$ with variations  of the Borel parameter $T^2$, where the $A$ and $B$ correspond to the $\Pi_{\rm RC}(T^2)$ and $\overline{\Pi}_{\rm RC}(T^2)$, respectively.  }\label{residue-Zc-MM}
\end{figure}

We take into account  the uncertainties  of the input   parameters,
and obtain the values of  the mass and pole residue, which are shown explicitly in Figs.\ref{mass-Zc-MM}-\ref{residue-Zc-MM},
\begin{eqnarray}
M_Z&=&3.85\pm0.09\, {\rm GeV}\, ,\nonumber\\
\lambda_Z&=&(1.35\pm0.24)\times 10^{-2}\,{\rm GeV}^5\,\,\,{\rm with} \,\,\,\,{\Pi_{\rm RC}(T^2)} \, ,\\
M_Z&=&3.89\pm0.08\, {\rm GeV}\, ,\nonumber\\
\lambda_Z&=&(1.66\pm0.25)\times 10^{-2}\,{\rm GeV}^5\,\,\,{\rm with} \,\,\,\,{\overline{\Pi}_{\rm RC}(T^2)} \, .
\end{eqnarray}
Although the values of the mass are both in consistent with the experimental data $M_{Z_c(3900)}=(3887.2\pm2.3)\,\rm{ MeV}$ from the Particle Data Group  \cite{PDG}, the value $M_Z=3.89\pm0.08\, {\rm GeV}$ with the $\overline{\Pi}_{\rm RC}(T^2)$ is better.
In Figs.\ref{mass-Zc-MM}-\ref{residue-Zc-MM}, we plot the mass and pole residue with variations of the Borel parameter $T^2$ at much larger range than the Borel window.

If we switch off the two-particle scattering state contributions $\Pi_{\rm RC}(T^2)$, we can obtain the values,
\begin{eqnarray}
M_Z&=&3.90\pm0.08\, {\rm GeV}\, ,\nonumber\\
\lambda_Z&=&(2.09\pm0.33)\times 10^{-2}\,{\rm GeV}^5 \, .
\end{eqnarray}
Compared with the value  $M_Z=3.90\pm0.08\, {\rm GeV}$ from the single-pole approximation \cite{Wang-Hidden-charm}, the values $M_Z=3.85\pm0.09\, {\rm GeV}$
and $3.89\pm0.08\, {\rm GeV}$ from the QCD sum rules where
the two-particle scattering state contributions included are slightly smaller, see Fig.\ref{mass-Zc-MM}.  While the values of the pole residue  $\lambda_Z=(1.35\pm0.24)\times 10^{-2}\,{\rm GeV}^5$ and
$(1.66\pm0.25)\times 10^{-2}\,{\rm GeV}^5$ are much smaller than the value  $(2.09\pm0.33)\times 10^{-2}\,{\rm GeV}^5 $ from the single-pole approximation, see Fig.\ref{residue-Zc-MM}. In all the QCD sum rules, there appear Borel platforms in the Bore window $T^2=(2.7-3.1)\,\rm{GeV}^2$. Moveover, the energy scale formula
$\mu=\sqrt{M^2_{X/Y/Z}-(2{\mathbb{M}}_c)^2}$ with the updated effective $c$-quark mass ${\mathbb{M}}_c=1.82\,\rm{GeV}$ \cite{Wang-tetra-formula,WangZG-eff-Mc} is satisfied for the QCD sum rules with or without the two-particle scattering state contributions $\overline{\Pi}_{\rm RC}(T^2)$, as the $\Pi_{\rm RC}(T^2)$  overestimates the two-particle scattering state contributions.
All in all, the pole contribution of the $Z_c(3900)$ is necessary, the current $J_\mu(x)$ couples potentially to the $Z_c(3900)$.

 Up to now, only the decays $Z_c(3900)\to J/\psi \pi$, $D\bar{D}^*$, $\eta_c\rho$ have been observed, and the branching fractions have not been determined yet. In Ref.\cite{WangZhang-Solid}, we choose the axialvector current $J_\mu(x)$ to interpolate the $Z_c(3900)$, and study the hadronic coupling  constants $G_{Z_cJ/\psi\pi}$, $G_{Z_c\eta_c\rho}$, $G_{Z_cD \bar{D}^{*}}$ with the QCD sum rules based on solid quark-hadron duality,  then study the two-body strong decays
and obtain the partial decay widths
$\Gamma(Z_c^+(3900)\to J/\psi\pi^+)=25.8\pm 9.6\,\rm{MeV}$, $\Gamma(Z_c^+(3900)\to\eta_c\rho^+)=27.9\pm 20.1 \,\rm{MeV}$,
$\Gamma(Z_c^+(3900)\to D^+ \bar{D}^{*0})=0.22\pm 0.07\,\rm{MeV}$, $\Gamma(Z_c^+(3900)\to \bar{D}^0 D^{*+})=0.23\pm 0.07\,\rm{MeV}$,
and the total width
$\Gamma_{Z_c}=54.2\pm29.8\,\rm{MeV}$, which is compatible with the experimental data \cite{BES3900,PDG}, and  also supports assigning the $Z_c^\pm(3900)$ to be  the diquark-antidiquark  type axialvector  tetraquark  state. Direct calculations of other partial decay widths with the QCD sum rules are also valuable  to make a more robust conclusion, this is our next work.

\subsection{Two-particle contributions only}
Again we choose the  continuum threshold parameter $\sqrt{s_0}=4.45\pm 0.10\,\rm{GeV}$ and the optimal energy scale $\mu=1.4\,\rm{GeV}$, the pole contribution  is
${\rm PC}=(43-66)\%$
at the Borel window $T^2=(2.7-3.1)\,\rm{GeV}$.  As there are only the two-particle scattering state contributions, the QCD sum rules are reduced to two
independent QCD sum rules in Eq.\eqref{QCD-Two-Particle} and Eq.\eqref{QCD-Two-Particle-derive}.

\begin{figure}
 \centering
 \includegraphics[totalheight=6cm,width=7cm]{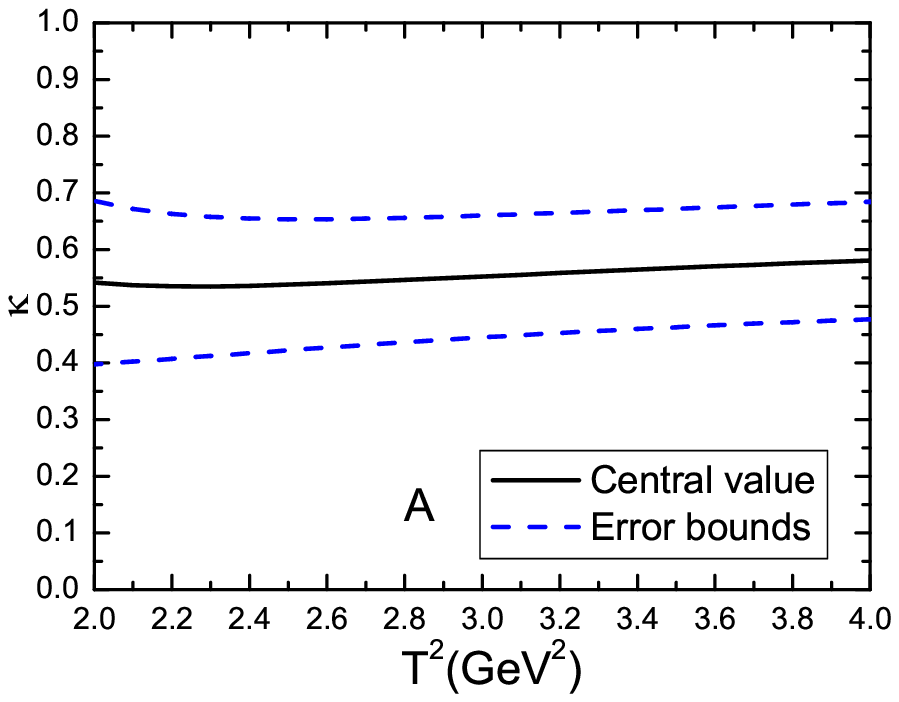}
 \includegraphics[totalheight=6cm,width=7cm]{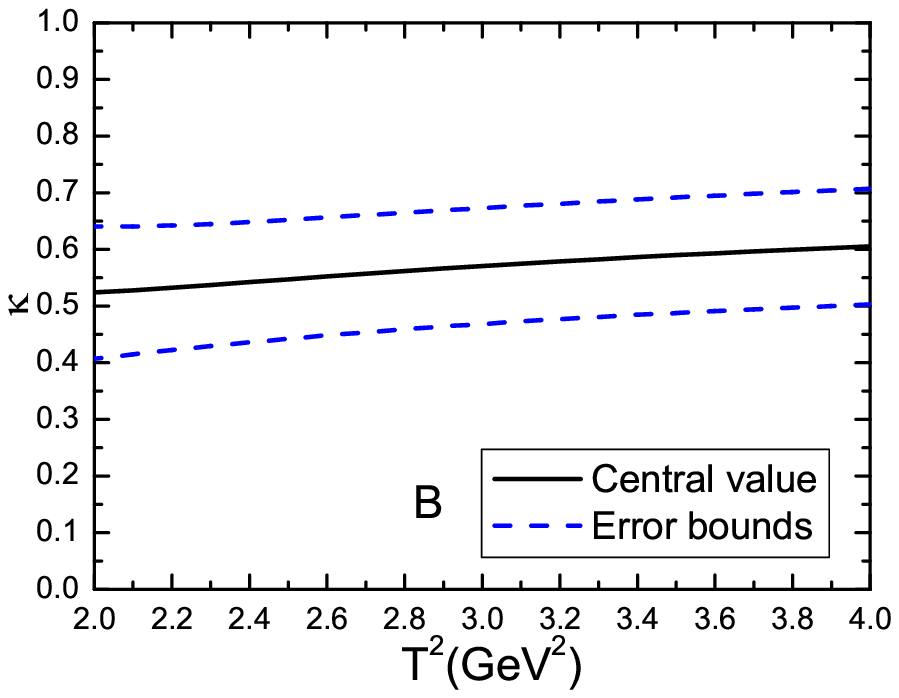}
 \caption{ The $\kappa$   with variations  of the Borel parameter $T^2$, where the $A$ and $B$ correspond to the QCD sum rules in Eq.\eqref{QCD-Two-Particle} and Eq.\eqref{QCD-Two-Particle-derive}, respectively.  }\label{kappa-fig}
\end{figure}

We take into account  the uncertainties  of the input   parameters,
and obtain the values of  the $\kappa$, which are shown explicitly in Fig.\ref{kappa-fig},
\begin{eqnarray}
\kappa&=&0.55\pm0.11\,  \,\,\,{\rm from} \,\,\,\,{\rm Eq.}\eqref{QCD-Two-Particle} \, ,\nonumber\\
&=&0.57\pm0.11\,  \,\,\,{\rm from} \,\,\,\,{\rm Eq.}\eqref{QCD-Two-Particle-derive}\, ,
\end{eqnarray}
in the Borel window. If we take the replacement $\Pi_{\rm RC}(T^2) \to \overline{\Pi}_{\rm RC}(T^2) $ to subtract the perturbative corrections, we obtain the
corresponding values $\overline{\kappa}$,
 \begin{eqnarray}
\overline{\kappa}&=&0.39\pm0.08\,  \,\,\,{\rm from} \,\,\,\,{\rm Eq.}\eqref{QCD-Two-Particle} \, ,\nonumber\\
&=&0.40\pm0.08\,  \,\,\,{\rm from} \,\,\,\,{\rm Eq.}\eqref{QCD-Two-Particle-derive}\, .
\end{eqnarray}

The values $\kappa$, $\overline{\kappa}\ll 1$, the two-particle scattering state contributions cannot saturate  the QCD sum rules at the hadron side.
If we  insist on performing the Fierz rearrangement,
\begin{eqnarray}
J_{\mu} &=&\frac{1}{2\sqrt{2}}\Big\{iJ^1_\mu-iJ^2_\mu- iJ^3_\mu+iJ^4_\mu+J^5_\mu-J^6_\mu    - i J^7_\mu+iJ^8_\mu   \,\Big\} \, ,
\end{eqnarray}
we should take into account the possible tetraquark  molecular states, such as $\eta_c\rho$, $\pi J/\psi$, $\cdots$,
\begin{eqnarray}
\langle 0|J^1_\mu(0,0)|\eta_c\rho(p)\rangle&=&\lambda_{\eta_c\rho,M}\,\varepsilon_{\mu}\, ,
\nonumber\\
\langle 0|J^2_\mu(0,0)|\pi J/\psi(p)\rangle&=&\lambda_{\pi J/\psi,M}\,\varepsilon_{\mu}\, ,
\end{eqnarray}
$\cdots$, where the $\lambda_{\eta_c\rho,M}$, $\lambda_{\pi J/\psi,M}$, $\cdots$
 are the pole residues, the $\varepsilon_\mu$ are the polarization  vectors.
  All in all, the two-particle  scattering state contributions cannot saturate  the QCD sum rules.
In Ref.\cite{WangZG-Landau}, we illustrate that the color-singlet-color-singlet type current couples potentially both to the tetraquark molecular states and two-meson scattering states.  The two-meson scattering states cannot saturate the QCD sum rules,  while
the tetraquark molecular states can saturate the QCD sum rules. We can take into account the two-meson scattering states reasonably by adding a finite width to the
tetraquark molecular states in the QCD sum rules for  the color-singlet-color-singlet type currents.
Furthermore, in the light-flavor sector,    Lee and Kochelev study  the two-pion  contributions in the QCD sum rules for the scalar meson $f_0(500)$  as the tetraquark state, and observe that the contributions of the order $\mathcal{O}(\alpha_s^k)$  with $k\leq1$ cannot be canceled out by the two-pion  scattering states \cite{Lee-0702-PRD}.

For the meson-meson type (or color-singlet-color-singlet type)  currents,  Lucha, Melikhov and Sazdjian assert that the disconnected Feynman diagrams at the order $\mathcal{O}(\alpha_s^k)$ with $k\leq1$ are exactly canceled out by    the two-meson scattering state  contributions  \cite{Chu-Sheng-PRD,Chu-Sheng-EPJC}, which lead to the two QCD sum rules shown in Eq.\eqref{QCD-Two-Particle} and Eq.\eqref{QCD-Two-Particle-derive}  in the present case. However, the small values $\kappa$, $\overline{\kappa}\ll 1$ reject those assertions. The tetraquark (molecular) states have to be taken into account  in the QCD sum rules, and they begin to receive contributions at the order $\mathcal{O}(\alpha_s^0)$.

\subsection{$Z_c(3900)$ only with the factorizable Feynman diagrams}
We choose the  continuum threshold parameter $\sqrt{s_0}=4.40\pm 0.10\,\rm{GeV}$ and the optimal energy scale $\mu=1.4\,\rm{GeV}$, the pole contribution  is
${\rm PC}=(40-63)\%$
at the Borel window $T^2=(2.7-3.1)\,\rm{GeV}$ with both the factorizable and nonfactorizable  Feynman diagrams.

\begin{figure}
 \centering
 \includegraphics[totalheight=6cm,width=8cm]{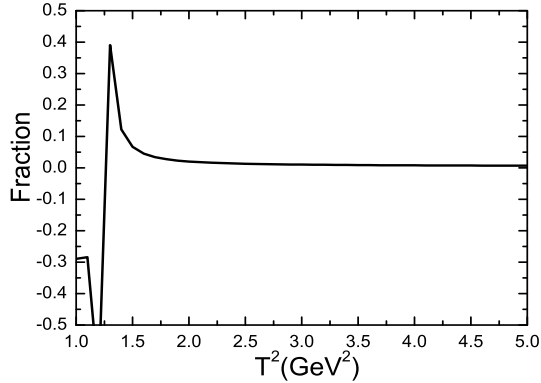}
 \caption{ The contribution from nonfactorizable  Feynman diagrams   with variations  of the Borel parameter $T^2$.  }\label{Non-factor-fraction}
\end{figure}

In Fig.\ref{Non-factor-fraction}, we plot the contribution from nonfactorizable  Feynman diagrams   with variations  of the Borel parameter $T^2$. From the figure, we can see that the nonfactorizable contribution is about $1\%$ at the Borel window and play a minor important role in the QCD sum rules. No stable QCD sum rules can be obtained in Eq.\eqref{QCD-fac-nonfac-two-par}.
The dominant  contributions come from the factorizable Feynman diagrams. The factorizable contributions consist of two colored  clusters,  a diquark cluster and an antidiquark cluster, the color confinement frustrates
tunnelling effects and leads to stable tetraquark state, which is consistent with the small width of the $Z_c(3900)$.

From the QCD sum rules in Eq.\eqref{QCD-fac-nonfac-Zc} and Eq.\eqref{QCD-fac-nonfac-Zc-derive}, we can obtain the values of the mass and pole residue,
\begin{eqnarray}
M_Z&=&3.90\pm0.08\, {\rm GeV}\, ,\nonumber\\
\lambda_Z&=&(2.09\pm0.33)\times 10^{-2}\,{\rm GeV}^5 \, ,
\end{eqnarray}
the nonfactorizable contributions can be neglected safely in the Borel window.

In {\it The Review of Particle Physics}, the $Z_c(3900)$ and $Z_c(3885)$ are taken as the same particle, and are named as $Z_c(3900)$ by the Particle Data Group \cite{PDG}. In fact, the $Z_c(3900)$ and $Z_c(3885)$
 have almost degenerated masses but quite  different decay widths, $\Gamma_{Z_c(3900)}=46 \pm 10 \pm 20\,\rm{MeV}$  and  $\Gamma_{Z_c(3885)}=24.8 \pm 3.3 \pm 11.0\,\rm{MeV}$  from the BESIII collaboration \cite{BES3900,BES-3885}. We can reproduce the mass of the $Z_c(3900)$ or $Z_c(3885)$ both in the scenarios of tetraquark  state and tetraquark molecular state \cite{WangHuangTao-3900,Molecular3900-ZhangJR,WangMolecule-3900,Wang-Hidden-charm}, if the $Z_c(3900)$ and $Z_c(3885)$ are  the same particle, it should have both the diquark-antiquark type tetraquark and meson-meson type molecule components. Without introducing the molecule component, it is difficult to take into account the  ratio $R_{exp}$,
  \begin{eqnarray}
  R_{exp} =\frac{\Gamma(Z_c(3885)\to D\bar{D}^*)}{\Gamma(Z_c(3900)\to J/\psi \pi)}=6.2 \pm 1.1 \pm 2.7 \, ,
 \end{eqnarray}
from  the BESIII collaboration \cite{BES-3885}. We can assign the $Z^+_c(3900)$ to be the  diquark-antidiquark type axialvector tetraquark state, and assign the $Z_c^+(3885)$
to be  the $D^+\bar{D}^{*0}+D^{*+}\bar{D}^0$ tetraquark molecular state  according to the predicted mass $3.89\pm 0.09\,\rm{GeV}$ from the QCD sum rules \cite{WangMolecule-3900}. If the $Z_c(3885)$ is the $D^+\bar{D}^{*0}+D^{*+}\bar{D}^0$ tetraquark molecular state, the decays to the final states  $D^+\bar{D}^{*0}$ and $D^{*+}\bar{D}^0$ take place through its component directly, it is easy to account for the large ratio $R_{exp}$.

\section{Conclusion}
In this article, we study the $Z_c(3900)$ with the QCD sum rules in details by including the contributions of the two-particle scattering states and nonlocal effects between the diquark and antidiquark constituents.
The two-particle scattering state contributions alone  cannot saturate  the QCD sum rules at the hadron side, we have to take into account the pole  contribution of the $Z_c(3900)$.
If we approximate the hadron side of the QCD sum rules with the  $Z_c(3900)$ plus two-particle scattering state contributions, we can obtain a mass which is consistent with the experimental data. In fact, we can saturate the QCD sum rules with or without the two-particle scattering state contributions, although different pole residues are  needed.
If there exists a repulsive barrier or spatial distance between the diquark and antidiquark constituents,  the Feynman diagrams can be divided into the factorizable and nonfactorizable diagrams. The factorizable contributions consist of two colored  clusters,  a diquark cluster and an antidiquark cluster, the color confinement frustrates
tunnelling effects and leads to a stable tetraquark state, which is consistent with the small width of the $Z_c(3900)$. The nonfactorizable Feynman diagrams correspond to the tunnelling effects, which play a minor important role in the QCD sum rules. If we equal the nonfactorizable contributions to the two-particle scattering state contributions, no stable QCD sum rules can be obtained. The factorizable contributions dominate  the QCD sum rules at the QCD side. The present conclusion is expected to apply to other QCD sum rules for the diquark-antidiquark type tetraquark states.
It is feasible to apply the QCD sum rules to study the tetraquark states, which  begin to receive contributions at the order   $\mathcal{O}(\alpha_s^0)$, not at the order $\mathcal{O}(\alpha_s^2)$.

\section*{Acknowledgements}
This  work is supported by National Natural Science Foundation, Grant Number  11775079.

\end{document}